\documentclass[acmsmall]{acmart}
\usepackage{subfigure} 
\usepackage{tabulary}
\usepackage{verbatim}
\usepackage{mathrsfs}
\usepackage{tcolorbox}
\usepackage{amsmath}

\usepackage{amssymb}
\usepackage{indentfirst}
\usepackage{booktabs}
\usepackage{colortbl}
\usepackage{threeparttable}

\makeatletter
\newcommand\figcaption{\def\@captype{figure}\caption}
\newcommand\tabcaption{\def\@captype{table}\caption}
\makeatother

\AtBeginDocument{%
  \providecommand\BibTeX{{%
    \normalfont B\kern-0.5em{\scshape i\kern-0.25em b}\kern-0.8em\TeX}}}

\setcopyright{acmcopyright}
\acmJournal{CSUR}
\acmYear{2022} \acmVolume{1} \acmNumber{1} \acmArticle{1} \acmMonth{1} \acmPrice{15.00}\acmDOI{10.1145/3544968}

\acmJournal{JACM}
\acmVolume{37}
\acmNumber{4}
\acmMonth{2}

\begin{document}

\title{Deep Learning for Android Malware Defenses: a Systematic Literature Review}

\graphicspath{{images/}}
\author{Yue Liu}
\email{yue.liu1@monash.edu}
\author{Chakkrit Tantithamthavorn}
\email{chakkrit@monash.edu}
\author{Li Li}
\email{Li.Li@monash.edu}
\affiliation{%
  \institution{Monash University}
  \city{Melbourne}
  \country{Australia}
}

\author{Yepang Liu}
\email{liuyp1@sustech.edu.cn}
\affiliation{%
  \institution{Southern University of Science and Technology}
  \city{Shenzhen}
  \country{China}
}

\renewcommand{\shortauthors}{Liu et al.}

\begin{abstract}
Malicious applications (particularly those targeting the Android platform) pose a serious threat to developers and end-users.
Numerous research efforts have been devoted to developing effective approaches to defend against Android malware. 
However, given the explosive growth of Android malware and the continuous advancement of malicious evasion technologies like obfuscation and reflection, Android malware defense approaches based on manual rules or traditional machine learning may not be effective.
In recent years, a dominant research field called deep learning (DL), which provides a powerful feature abstraction ability, has demonstrated a compelling and promising performance in a variety of areas, like natural language processing and computer vision.
To this end, employing deep learning techniques to thwart Android malware attacks has recently garnered considerable research attention.
Yet, no systematic literature review focusing on deep learning approaches for Android malware defenses exists.
In this paper, we conducted a systematic literature review to search and analyze how deep learning approaches have been applied in the context of malware defenses in the Android environment. 
As a result, a total of 132 studies covering the period 2014-2021 were identified. 
Our investigation reveals that, while the majority of these sources mainly consider DL-based Android malware detection, 53 primary studies (40.1\%) design defense approaches based on other scenarios.
This review also discusses research trends, research focuses, challenges, and future research directions in DL-based Android malware defenses.
\end{abstract}



\maketitle

\section{Introduction}
Android is one of the most popular smartphone operating systems (OS), having dominated more than 70\% of the mobile OS market share since October 2016, according to a Statista report~\cite{Statista}.
Due to its openness and popularity, Android has become one of the primary targets of cyber-attacks~\cite{faruki2014android}.
Developers may take advantage of crafted malicious applications to divulge mobile user privacy or perform other dangerous operations on users' mobiles, which is extremely harmful to mobile users.
On the other hand, there is a large scale of Android apps in the real world, with over 3 million Android apps available through the official store, Google Play.
Although Google is constantly upgrading its protection against malicious attacks and has developed Google Play Protection (GPP)~\cite{GPP}, it is not reliable and researchers have proved that crafted dangerous apps can easily bypass the GPP's detection~\cite{hutchinson2019we,li2015iccta,dong2018frauddroid,liu2020maddroid}. 
Apart from the official market, there are hundreds of unofficial and third-party markets, where the security of Android apps is highly unpredictable~\cite{li2019revisiting, li2017understanding, zhan2020automated, zhang2020empirical}. 
Therefore, it is a pressing demand to propose an available and reliable approach to defend against malware attacks on the Android platform. 

Android malware defenses are a critical research topic in computer security. 
Manually analyzing malware, by formulating corresponding rules and inspecting the behaviors and source code of suspicious Android apps, is a time-consuming process---i.e., it does not scale to a large amount of Android software. 
Besides, with malware techniques constantly evolving, manual malware analysis couldn't keep pace with the evolving attack strategies. 
In recent years, a large volume of research related to automatic Android malware analysis has been proposed, utilizing data mining and machine learning approaches to achieve acceptable malware detection performance. 
These approaches employ a series of machine learning algorithms (e.g., support vector machine, random forest) to build a prediction model based on feature vectors extracted from the Android application package (APK)~\cite{sahs2012machine,arp2014drebin,wu2012droidmat}.
However, traditional machine learning algorithms are limited in their ability to learn complicated representations in high-dimensional spaces~\cite{lecun2015deep}. 
In addition, the performance of machine learning models heavily relies on the training data, and these trained models are likely to become obsolete as the Android apps evolve and software engineering advances.
What's more, attackers continue to update their fraud techniques to bypass protection software as well as well-trained machine learning models in order to victimize users and businesses. 
In front of the increasing difficulty of Android malware defenses, it is 
non-trivial to construct a robust and transparent defense model or system only by traditional machine learning techniques~\cite{zhao2021impact}.

Deep learning has emerged as the dominant research field of machine learning over the last decade, with notable achievements in many domains like speech recognition~\cite{hinton2012deep,amodei2016deep} and image processing~\cite{szegedy2015going,xie2017aggregated}. 
In contrast to conventional machine learning techniques, feature extraction can be performed automatically when deep learning methods are fed with raw data.
Deep learning can learn feature representation from the inputted raw data with little prior knowledge, which is the key advantage of deep learning. 
In 2014, deep learning tools were applied to Android malware defenses and demonstrated superior performance~\cite{yuan2014droid}. 
Subsequently, an increasing number of researchers have developed Android malware defenses models or frameworks based on a variety of deep learning techniques. 
As a result, an up-to-date comprehensive survey of DL-based Android malware defenses is urgently required.

\begin{table}[t]
  \centering
  \scriptsize
  \caption{Summary of Related work}
    \begin{tabular}{lccp{8em}p{15.25em}r}
    \toprule
    \multicolumn{1}{c}{\textbf{Paper}} & \textbf{Ref. size} & \textbf{Newest Ref.} & \multicolumn{1}{c}{\textbf{Scope}} & \multicolumn{1}{c}{\textbf{Review Approach}} & \multicolumn{1}{p{6.75em}}{\textbf{Research trend analysis}} \\
    \midrule
    Alqahtani et al.~\cite{alqahtani2019survey} & 9     & 2019  & Malware detection & Informal & No \\
    Souri et al.~\cite{souri2018state} & 47    & 2018  & Malware detection & Informal & No \\
    Qiu et al.~\cite{qiu2020survey} & 46    & 2019  & Malware detection & Informal & No \\
    Naway et al.~\cite{naway2018review} & 25    & 2018  & Malware detection & Systematic search(method details not described) & No \\
    Liu et al.~\cite{liu2020review} & 113   & 2019  & Malware detection & Systematic search & No \\
    Wang et al.~\cite{wang2020review} & 54    & 2020  & Malware detection & Informal & No \\
    This work & 132   & 2021  & Malware defenses & Systematic search + snowballing + quality analysis & Yes \\
    \bottomrule
    \end{tabular}%
  \label{tab:relatedwork}
\end{table}%

The domain of Android malware defenses has been widely researched in recent years, and we present related contributions of other researchers in Table~\ref{tab:relatedwork}. 
Several early studies~\cite{faruki2014android,li2017static,tam2017evolution} have comprehensively reviewed Android malware techniques and traditional defensive approaches. 
With the wide use of advanced machine learning techniques, many researchers have reviewed relevant studies on Android malware defenses with machine learning or deep learning~\cite{alqahtani2019survey, souri2018state, qiu2020survey, naway2018review, wu2020systematical, wang2020review}. 
However, these previous works couldn't provide a complete picture of current research interests and trends on DL-based Android malware defenses though they analyze all possible available methods. 
First, these previous studies focus only on one aspect of Android malware defenses, using machine learning/deep learning techniques to detect Android malware (ML/DL-based malware detection), but neglect other critical aspects of using deep learning to prevent/defend against malicious behaviors (e.g., malware evolution, adversarial malware detection, deployment, malware families). 
While distinguishing malware from benign apps is critical, enhancing Android software security is not a straightforward binary classification task.
Indeed, it requires not only locating malicious applications but also comprehending malicious behaviors, to which many researchers have contributed. 
However, these research studies are overlooked from previous review work, making it difficult for future researchers to comprehend the state of the art of this research field. 
More importantly, these early surveys are not based on completed systematic approaches, and thus they could not provide a comprehensive overview of the research trends and open issues in this domain.
Thus, a number of unanswered questions remain regarding the development of deep learning-based Android malware defenses.
For example, the prior works still could not answer what are the state-of-the-art DL-based malware defense approaches (e.g., deep learning models and feature processing approaches) and what aspects require more research efforts in the future.
Furthermore, most previous works focused on relevant studies published before 2019. 
However, DL-based Android malware defenses have attracted significant research attention in recent two years, which means it is necessary to conclude the significant recent research achievements. 
As a result, this article fills the research gap in this field by conducting a systematic and organized literature review, summarizing previous research and presenting research trends on Android malware defenses related to deep learning.

This survey aims to shape the research area of using deep learning techniques to defend against Android malware, and position existing works and current progress.
Specifically, this paper makes the following contributions:
\begin{itemize}
\item We systematically collect and review 132 primary studies published between 2014 and 2021 on DL-based Android malware defenses.
\item We present a comprehensive qualitative and quantitative synthesis based on the collected studies. 
Our synthesis covers the following themes: research objectives, APK characterization, deep learning techniques, deployment, and model evaluation. 
\item We further enumerate current issues of the existing works from different aspects and provide recommendations based on findings to support further research in this domain.
\item We provide trend analysis to identify potential future trends for the research community.
\end{itemize} 

The remainder of this paper is structured as follows: Section 2 presents the review methodology used in this paper. Section 3 discusses the reviewed results and open issues for the proposed research questions. Section 4 and 5 discuss potential implications and possible threats to validity of this study respectively. Finally, Section 6 concludes the paper.

\section{Review Methodology}
In this paper, we followed the methodology suggested by Kitchenham~\cite{kitchenham2004procedures} to conduct a systematic review. 
The main steps of the Systematic Literature Review (SLR) can be summarized as follows: (1) planning the review and developing a review protocol, (2) identifying research questions, (3) designing search strategies, proposing exclusion criteria, (4) data extraction, and (5) data synthesis. 
The following subsections discuss the review protocol used in this paper.
Due to page limitations, we detailed the systematic review process and results online as supplementary materials~\footnote{https://github.com/yueyueL/DL-based-Android-Malware-Defenses-review}.

\subsection{Research Question}
In this paper, we seek to investigate the following research questions:
\begin{itemize}
    \item \textbf{RQ1}: What are the research objectives of the DL-based Android malware defense solutions?
    \item \textbf{RQ2}: What approaches have been developed for malware defenses? 
        \begin{itemize}
        \item \textbf{RQ2.1}: How are features processed for model training?
        \item \textbf{RQ2.2}: What deep learning architectures are used?
        \item \textbf{RQ2.3}: How are DL-based Android malware defenses approaches deployed in practice? 
        \item \textbf{RQ2.4}: How are DL-based Android malware defenses approaches evaluated? 
        \end{itemize}
    \item \textbf{RQ3}: What are the emerging and potential research trends for DL-based Android malware defenses? 
\end{itemize}

\subsection{Search Strategy}
\label{sec:searchstrategy}
After identifying the research questions, the next step is searching for relevant primary studies. 
To this end, five popular digital libraries, including IEEE, ACM Digital Library, Springer, Science Direct, and Wiley Online Library, are identified and the searching string is constructed based on the proposed searching items proposed in Table~\ref{searchkeywords}. 
To ensure that we did not overlook any significant relevant work, we conducted further searching processes on two of the most popular research citation engines, including Web of Knowledge\footnote{https://webofknowledge.com} and Google Scholar.\footnote{https://scholar.google.com}
In addition, we also performed a lightweight backward snowballing \cite{kitchenham2007guidelines}, which means that we only carried out snowballing once, before we identified the final review list.

\begin{table*}
  \caption{Search Keywords}
  \label{searchkeywords}
  \scriptsize
  \begin{tabular}{p{1.5cm}|p{11cm}}
    \toprule
    \textbf{Group} & \textbf{Keywords}\\
    \midrule
    1& Android; Mobile; Smartphone*; Phone* \\  \hline
    2& Malware; Malicious; Malice\\ \hline
    3& "Deep learning"; "Deep neural network*"; DNN; "Convolutional neural network*"; CNN; "Deep belief network*"; DBN; "Recurrent neural network*"; RNN; "Long short-term memory"; LSTM\\
    \bottomrule
  \end{tabular}
    \begin{tablenotes}    
        \footnotesize              
        \item Note: * means the plural form. For example, "Phone*" refers to "Phone" or "Phones".
      \end{tablenotes}  
\end{table*}

\subsection{Data Selection Process}
\label{sec:Selectionstrategy}
\begin{figure}[t]
  \centering
  \subfigure[Publication count over time]{ \includegraphics[width=7.2cm]{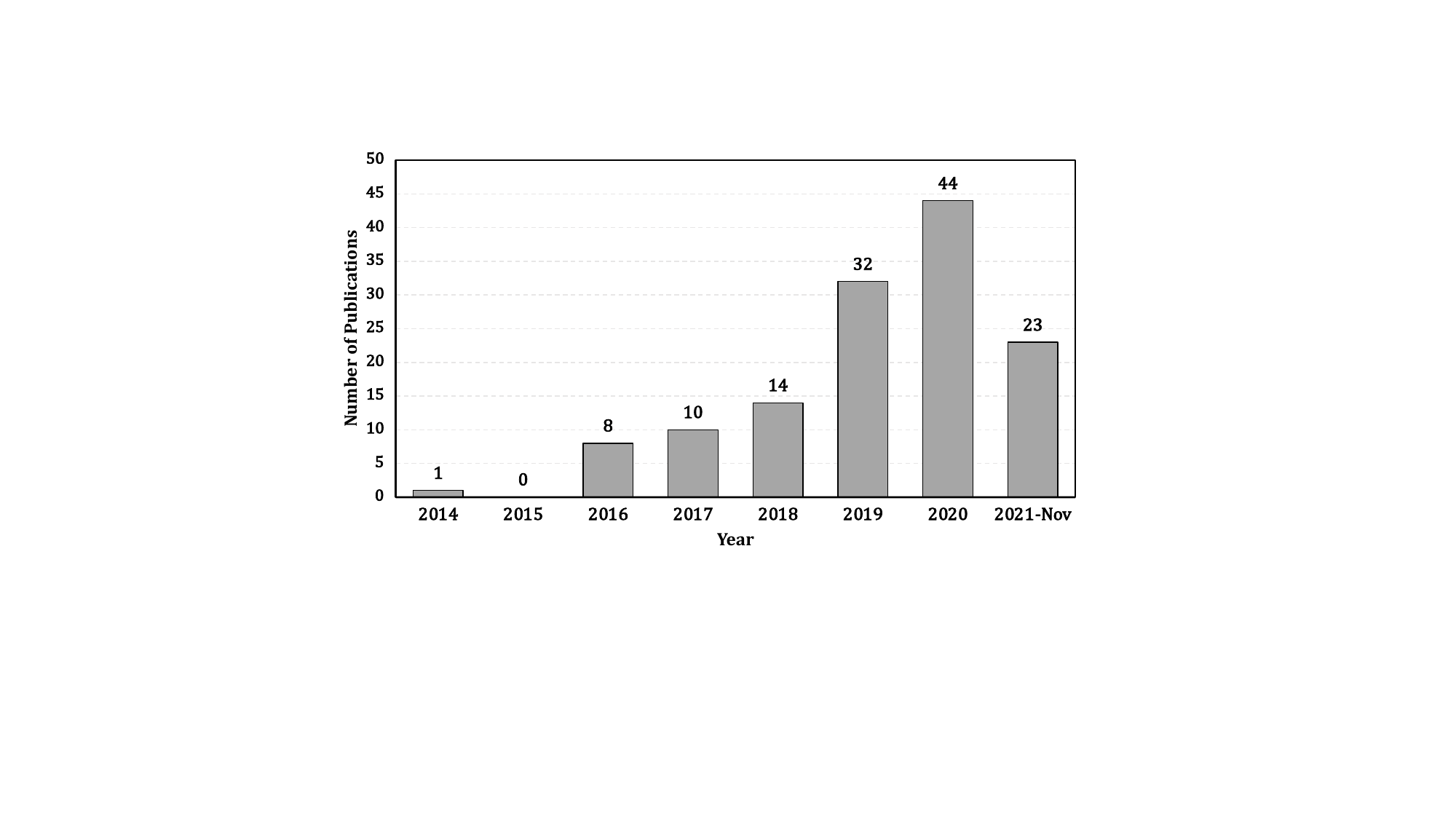}}
  \subfigure[Word cloud of all major venue names of the sources]{\includegraphics[width=6.5cm]{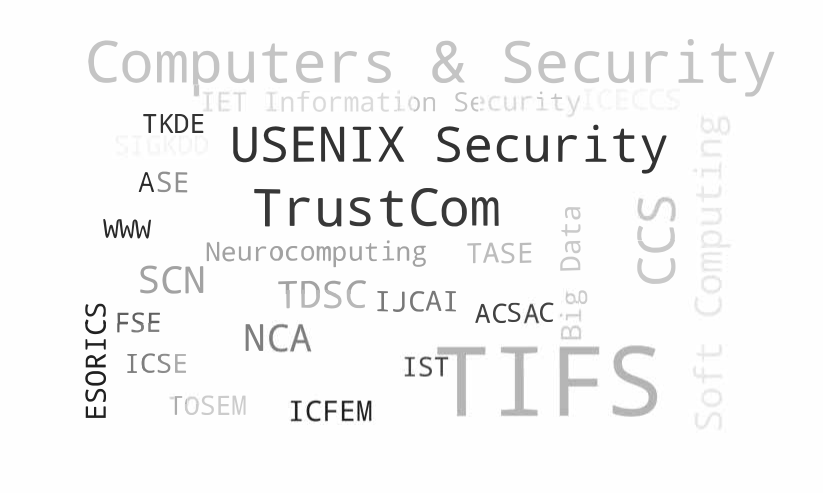}}
  \caption{Summary of the examined primary studies}
  \label{fig:ResultAnalysis}
\end{figure}

Only those studies related to deep learning-based Android malware defenses should be considered for further review; therefore, any primary studies that meets any of the proposed exclusion criteria would be deemed irrelevant and
would be excluded from the preliminary result set. 
On the other hand, obtaining all relevant studies doesn't guarantee that we are able to identify the final list of papers, as it is impossible that the quality of all selected studies is desirable. 
For this reason, we defined a quality appraisal criterion and evaluated the quality of each paper by reading its full text. 
The complete list of exclusion criteria and quality appraisal criterion is available at our online supplementary materials. 
After these steps, we finally obtained 132 primary studies.
Table~\ref{tab:searchresults} and in Figure~\ref{fig:ResultAnalysis} provided a summary for our examined papers.

Fig.~\ref{fig:ResultAnalysis}(a) shows the distribution of the amount of chosen studies over time. 
Intuitively, the number of publications related to DL-based Android malware defenses has seen a continued increase since 2014. 
Although we only included the public articles before November 30, 2021, in this review, the number of selected publications in 2021 is still large. 
These facts demonstrate that the field of Android malware defenses using deep learning is attracting growing attention, illustrating the critical need for systematic and comprehensive review work to summarize the prior work and current research trends.
 
On the other hand, we examined the distribution of venue domain and type for these 132 articles respectively. 
The results showed that over 35\% of primary studies are from Security (SEC) venues, accounting for the most proportion. 
Both the proportion of Artificial Intelligence (AI) and Software Engineering/Programming Languages (SE/PL) is more than 10\%. 
As for the type of venues, we found the percentage of collected studies published in conferences and journals is quite close, at about 50\%. 
In addition, we counted the frequency of all major venues where our selected studies were published (see Fig.~\ref{fig:ResultAnalysis}(b)). 
The results indicated that these primary studies were mainly collected at top venues, especially the venues in SEC domain (e.g., CCS, USENIX Security, TIFS) and more and more relevant studies have started to be presented in top venues in SE domain recently (e.g., ICSE, ASE, and FSE).

\begin{table}[t]
\centering
\caption{Summary of the process of data search and selection}
  \small
  \begin{tabular}{lcc}
    \toprule
    \textbf{Data source} &\textbf{Number of search results}&\textbf{Number of Candidate studies}\\
    && \textbf{(After selection)} \\
    \midrule
    IEEE &201&108\\
    ACM&1182&35\\
    Springer&2404&36\\
    Science Direct&1031&19\\
    Wiley&457&8\\ \hline
    \multicolumn{2}{r}{Merge} & 206 \\
    \multicolumn{2}{r}{Further Searching} & 328 \\
    \multicolumn{2}{r}{After Quality Assessment} & 132 \\
    \multicolumn{2}{r}{After Backward Snowballing} & 132 \\
    \multicolumn{2}{r}{Final result} & 132 \\
  \bottomrule
\end{tabular}
  \label{tab:searchresults}
\end{table}%

\section{Results Analysis}
\label{sec: resultsanalysis}

In order to answer the research questions presented in Section~\ref{sec:Selectionstrategy}, we conducted a detailed review of the selected primary studies. 
Section~\ref{sec:malwareobjectives} discusses the analysis results for RQ1; Section~\ref{sec:Analysismethods},~\ref{sec:deeplearningtech},~\ref{sec:deploymentofanalysis}, and ~\ref{sec:PerformanceEvaluation} presents the results for RQ2.1, RQ2.2, RQ2.3, and RQ2.4 respectively; while Section~\ref{sec:trendanalysis} presents the results of RQ3. 
To help our fellow researchers better understand the details for each primary study, we uploaded a detailed table in our online supplementary materials.

\subsection{Malware Defenses Objectives}
\label{sec:malwareobjectives}
\begin{figure}[t]
  \centering
  \includegraphics[width=1\textwidth]{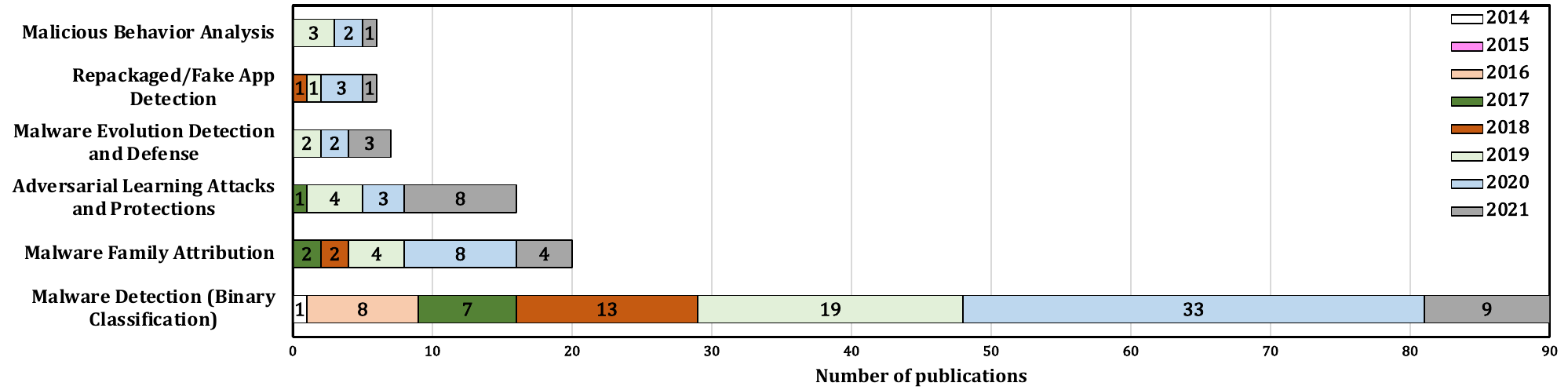}
  \caption{Summary of the primary studies by research objectives. Some primary papers contain multiple research objectives, making the sum of percentages more than 100\%.}
  \label{fig:RQ1trend}
\end{figure}

Deep learning techniques have been applied to various aspects of malware defenses to protect mobile users from severe malware attacks.
After discussing among all authors and drawing on the classification scheme used in previous surveys by Faruki et al.~\cite{faruki2014android} and Ucci et al.~\cite{ucci2019survey}, we classify reviewed studies into the following categories: malware detection (binary classification), malware family attribution, repackaged/fake app detection, adversarial learning attacks and protections, malware evolution detection and defense, and malicious behavior analysis. 
Fig.~\ref{fig:RQ1trend} depicts the statistical trends of research objectives for the sources.

\textbf{Malware Detection (Binary Classification).}
As shown in Fig.~\ref{fig:RQ1trend}, malware detection (binary classification), which determines whether a given application is malicious or benign, receives the most research attention (68\%) and the increasing trend is expected to continue. 
This result is not surprising given that the most urgent task at the moment is to protect mobile users from malicious attacks by automatically distinguishing malware from goodware, which is why many previous surveys have primarily focused on this research topic.
Droid-Sec~\cite{yuan2014droid} is the first attempt to detect Android malware using deep learning-based methods.
The methodology of Droid-Sec can be summarized as three steps: (1) Android applications collection and labeling, (2) feature extraction and characterization, (3) deep learning models training and evaluation. 
The empirical results of Droid-Sec have demonstrated that deep learning techniques are much more effective for malware detection compared with traditional machine learning techniques like Support Vector Machine(SVM).
In fact, most primary studies related to DL-based malware detection usually follow a similar methodology with Droid-Sec but explore the applicability and effectiveness of different state-of-arts deep learning techniques in more complex scenarios, which is consistent with previous literature~\cite{qiu2020survey}. 

\textbf{Malware Family Attribution.}
Another important aspect of Android malware defenses is malware family attribution.
Fig.~\ref{fig:RQ1trend} shows that 20 reviewed articles (15\%) are specialized for identifying Android malware families.
Given the growing number of malware variants, malware can be categorized into certain categories that are associated with different malicious objectives and behaviors, like the Adware family that displays unwanted advertisements to mobile users.
In contrast to malware detection (binary classification), malware family attribution identifies which family a malware sample belongs to.
Most primary studies like~\cite{zhiwu2019android} and ~\cite{sun2019android} employ multi-class classification approaches to identify existing or old malware families.
As a large number of new malware variants are created, Qiu et al.~\cite{qiu2019a3cm} proposed deep learning-based approaches to detect zero-day malware families.

\textbf{Repackaged/Fake App Detection.}
In 5\% of sources, deep learning-based repackaged/fake app detection is investigated.
Attackers can unpack an existing malicious/benign application, modify its contents and repackage it, depriving app developers of revenue and contributing to the spread of malware on mobile devices~\cite{li2019rebooting}. 
For this reason, identifying repackaged or fake applications and analyzing the behaviors of variants is also critical. 
For example, in order to locate counterfeit mobile applications in application markets, Ullah et al. \cite{ullah2019detection} and Karunanayake et al. \cite{karunanayake2020multi} propose DL-based Fake app detectors to prevent the publishing of fake apps in app stores.

\textbf{Adversarial Learning Attacks and Protections.}
Fig.~\ref{fig:RQ1trend} shows that 16 primary studies (12\%) focus on adversarial learning attacks and protections on DL-based malware defenses. 
Despite the fact that numerous research studies have demonstrated that deep learning models provide promisingly high performance to identify malware, these models have been shown to be particularly vulnerable to well-designed adversarial attacks~\cite{kurakin2016adversarial,yuan2019adversarial}.
Adversarial attackers could inject a small but intentional perturbation to create adversarial examples, causing the trained models to misclassify adversarial examples.
For example, Chen et al.~\cite{chen2019android} performed adversarial attacks on DNN-based malware detection models, decreasing the accuracy from over 90\% to 0\%. 
Consequently, there is a corresponding increase in the attention dedicated to adversarial attacks against malware defense models, as shown in Fig.~\ref{fig:RQ1trend}. 
Depending on when the attacks occur, adversarial attacks are split into two main categories: evasion attacks for testing samples and poisoning attacks for training samples. 
With respect to the two types of adversarial attacks, the majority of sources (14 studies, 87\%\%) discuss evasion attacks and protections for DL-based Android malware defense models, and conversely, only two recent studies focus on poisoning attacks~\cite{li2021backdoor,severi2021explanation}. 
We discuss more details about this topic in Section~\ref{sec:adversarial_trends}.

\textbf{Malware Evolution Detection and Defense.}
With regard to the malware evolution problem, Fig.~\ref{fig:RQ1trend} indicates that only seven papers (5\%) attempt to develop solutions for malware evolution, but it is remarkable that all seven papers were published within the last three years.
Due to the rapid evolution of mobile malware and the emergence of new variants and families, the performance of DL-based malware defenses models decays significantly over time.
Pendlebury et al.~\cite{pendlebury2019tesseract} revealed that the detection performance of deep learning-based classifiers decreases drastically from almost 90\% to below 30\% for future malware samples. 
Thus, model retraining and active learning are applied to reverse and improve aged models by Pendlebury et al.~\cite{pendlebury2019tesseract}.
However, the underlying models are still incapable of distinguishing evolved malware in this manner, as they still rely on humans to determine when models should be retrained.
In the light of this issue, recent studies~\cite{yang2021cade, xu2020sdac, fan2021heterogeneous, zhang2020enhancing, li2021can, lei2019evedroid} introduce a variety of approaches to slow down the aging of malware defense models, which are further discussed in Section~\ref{sec:malware_evolution_trend}. 

\textbf{Malicious Behavior Analysis.}
There are six primary studies (5\%) related to malicious behavior analysis in collected studies.
Malicious behavior analysis aims at identifying or assessing risk behaviors in unknown applications. 
As for Android malware, malicious behaviors have diverse types, and a malicious application often performs more than one malicious behavior, increasing the difficulty of analysis. 
In addition, malicious applications may utilize code obfuscation and dynamic payload to conceal malicious behaviors.
Hence, it is a relatively challenging research topic to investigate. 
In order to prevent malicious activities while apps are running, Gronat et al.~\cite{gronat2019maxnet} and Lorenzo et al.~\cite{de2020visualizing} employ recurrent neural networks to visualize potential risks for Android malware samples. 
For Android malware, performing malicious behaviors requires using dangerous semantic features such as permissions and API calls related to users' privacy. 
To assist mobile users in determining the security risk before installing unknown applications or granting permissions, some researchers examine the consistency between risk permissions and metadata-based features of apps, like descriptions~\cite{feng2019ac,feichtner2020understanding} or icon widgets~\cite{xi2019deepintent}.


~\\
\noindent \textbf{DISCUSSION.}  
Despite the rapidly growing number of research studies on deep learning for Android malware defenses, it appears that previous research studies focus on relatively simple application scenarios.
More than half of the sources focus on malware detection through various deep learning strategies.
Additionally, most of these existing studies focus on improving malware detection performance through the use of various advanced deep learning techniques and demonstrate that the newly proposed models outperform prior models on their own experimental datasets.
It is noteworthy that an increasing number of recent studies have started to address specific issues to better apply DL-based malware detection models in practice (e.g., on-device malware detection~\cite{feng2019mobidroid,feng2020performance}, explainable malware detection~\cite{zhu2019transparent, wu2020android}, malware detection on imbalanced data~\cite{bai2020unsuccessful,oak2019malware}). 
However, the number of relevant studies remains small.
How to improve the robustness, effectiveness, stability, and reliability of malware detectors with the help of deep learning is an open issue for future researchers. 


Compared with Android malware detection, the number of literature focusing on other research objectives is relatively small, requiring further in-depth research. 
Taking malware behavior analysis as an example, defining specific malicious behaviors and associating them with the raw code of Android APKs remain challenging issues.
Thus, these research objectives require more works to integrate domain knowledge and provide fundamental theoretical construction. 
Except that, while this review categorizes the existing literature's research objectives into six categories, the scope of Android malware defenses is actually much broader. 
Therefore, future research should not be limited to these six categories, but should instead propose Android malware defense approaches that leverage advanced deep learning techniques in more new application scenarios.

\begin{tcolorbox}
    \textbf{RQ1 What are the research objectives of the DL-based Android malware defense solutions? } 
    \begin{itemize}
        \item The main objective is still malware detection (binary classification) using deep learning techniques;
        \item On the whole, 53 primary studies focus on other research topics like malware family attribution and adversarial attacks, and the number is not small and cannot be neglected;
        \item At the beginning, researchers only focused on the field of malware detection, but in recent years, an increasing number of primary studies have applied deep learning to analyze Android malware in more complex scenarios.
    \end{itemize}
\end{tcolorbox}

\subsection{APK Characterization}
\label{sec:Analysismethods}

As a response to RQ2.1, this section discusses the APK feature processing approaches used in the collected studies.
Each Android application is packaged as an APK file, a zip archive that primarily contains the app's manifest and bytecode.
Before being fed into deep learning models, the collected Android APK data needs to be transformed into a formalized representation compatible with deep learning models.
These research studies usually process APK files using reverse-engineering tools (\textbf{program analysis approaches}) and then various raw characteristics (\textbf{feature categories}) are extracted.
After that, \textbf{feature encoding approaches} are utilized to perform further feature embedding operations on the raw information extracted from applications. 
To gain a better understanding of APK characterization mechanisms in DL-based Android malware defenses, we discuss the reviewed results from three perspectives, including program analysis approaches, feature categories, and feature encoding approaches.

\begin{figure}[t]
  \centering
  \subfigure[Trends of program analysis approaches] {\includegraphics[width=3.5in]{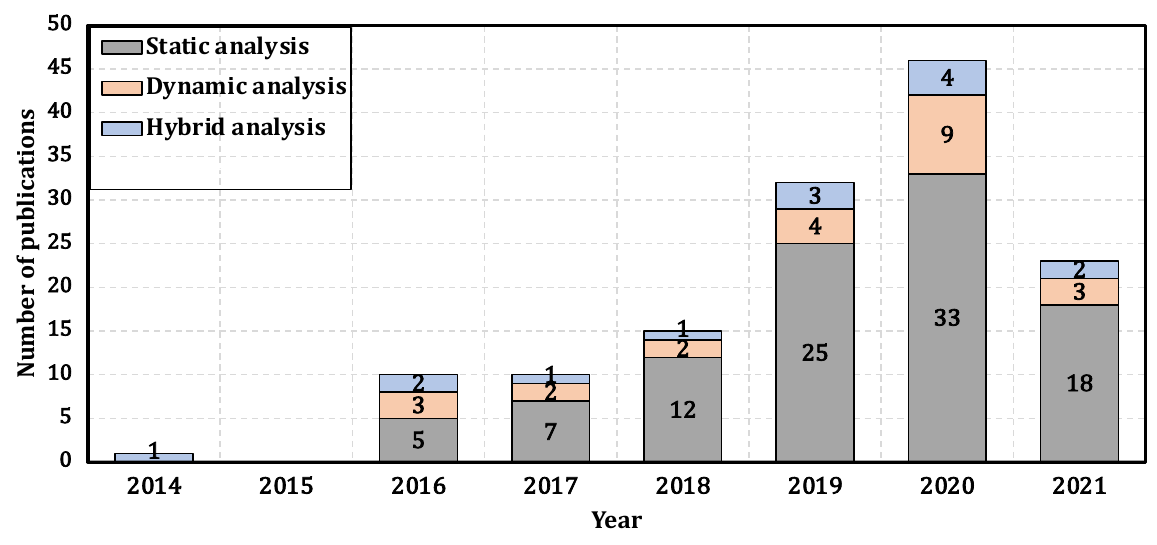}}
  \subfigure[Overall distribution]{\includegraphics[width=1.3in]{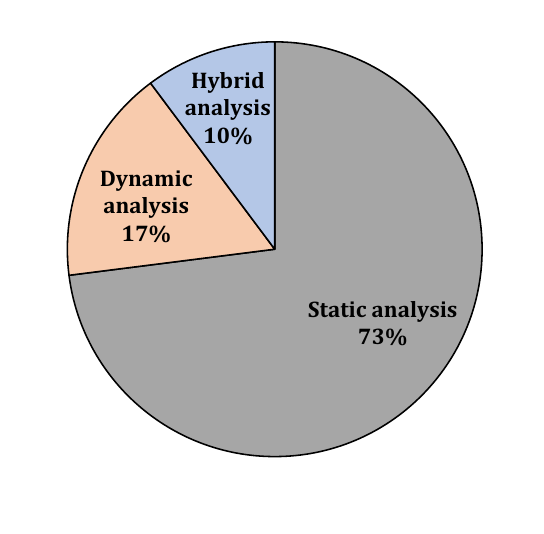}}
  \caption{Summary of the primary studies by program analysis approaches}
  \label{fig:RQ2programanalysis}
\end{figure}

\subsubsection{Program analysis approaches}
As shown in Fig.~\ref{fig:RQ2programanalysis}, program analysis approaches to extract raw features from Android APKs can be categorized into three types: static analysis, dynamic analysis, and hybrid analysis.

\textbf{Static Analysis.}
Fig.~\ref{fig:RQ2programanalysis} presents that the majority of sources (73\%) extract raw features using static analysis approaches. 
Reverse-engineering tools such as Androguard~\cite{Androguard} and APKtool~\cite{APKtool} are required to disassemble and/or decompile Android APK.
The raw information extracted from the APK files is used for further analysis of malicious applications.
The extracted information is diverse.
Raw binary code and opcode sequence can be fed directly to DL models~\cite{sun2019android, zhu2019fsnet, he2019malware, hsien2018r2}.
Aside from that, high-level semantic features like API calls and permissions are also widely used~\cite{su2016deep, grosse2017adversarial, kim2018multimodal, zhang2019familial}.

\textbf{Dynamic Analysis.}
Only 17\% of primary studies use dynamic analysis approaches to collect raw features from Android APK files.
This finding is not surprising given that dynamic analysis requires executing apps in a protected environment and dynamic analysis can only provide a partial picture of applications (i.e., it is challenging to cover all code)~\cite{faruki2014android, li2017static}.
However, dynamic analysis works by running samples to examine the runtime behaviors and system metrics of Android applications, which is more resilient to malware evasion techniques like obfuscation~\cite{hou2016deep4maldroid}.
Representative dynamic analysis tools include TaintDroid~\cite{enck2014taintdroid}, CopperDroid~\cite{tam2015copperdroid}, etc. 
Dynamic features are obtained by dynamically executing collected app samples in a controlled environment, such as an Android emulator or a real mobile device.
Thirteen primary studies employ emulators such as Genymotion to monitor the application's dynamic behaviors.
However, various anti-emulator techniques are developed to conceal malicious activities.
Thus, we also discovered that there are seven primary studies focusing on dynamic analysis on real mobile devices.
For example, Alzaylaee et al.~\cite{alzaylaee2017emulator} demonstrated that on-device dynamic analysis performed much better than on-simulator dynamic analysis concerning stability and detecting ability.

\textbf{Hybrid Analysis.}
Fig.~\ref{fig:RQ2programanalysis} presents that 10\% of primary studies involve hybrid program analysis (which combines static and dynamic analysis). 
Static program analysis has the advantage of providing full code coverage at a lower computational cost but it is vulnerable to evasion techniques like obfuscation, while dynamic program analysis allows for the analysis of run-time behaviors in a controlled environment but the code coverage may be limited~\cite{ vinayakumar2018detecting, chaulagain2020hybrid}.
Although hybrid analysis leverage the complementary strengths of both types of program analyses, it is still computationally intensive, which may explain why the number of related studies is small.

\begin{figure}[t]
  \centering
  \includegraphics[width=1\textwidth]{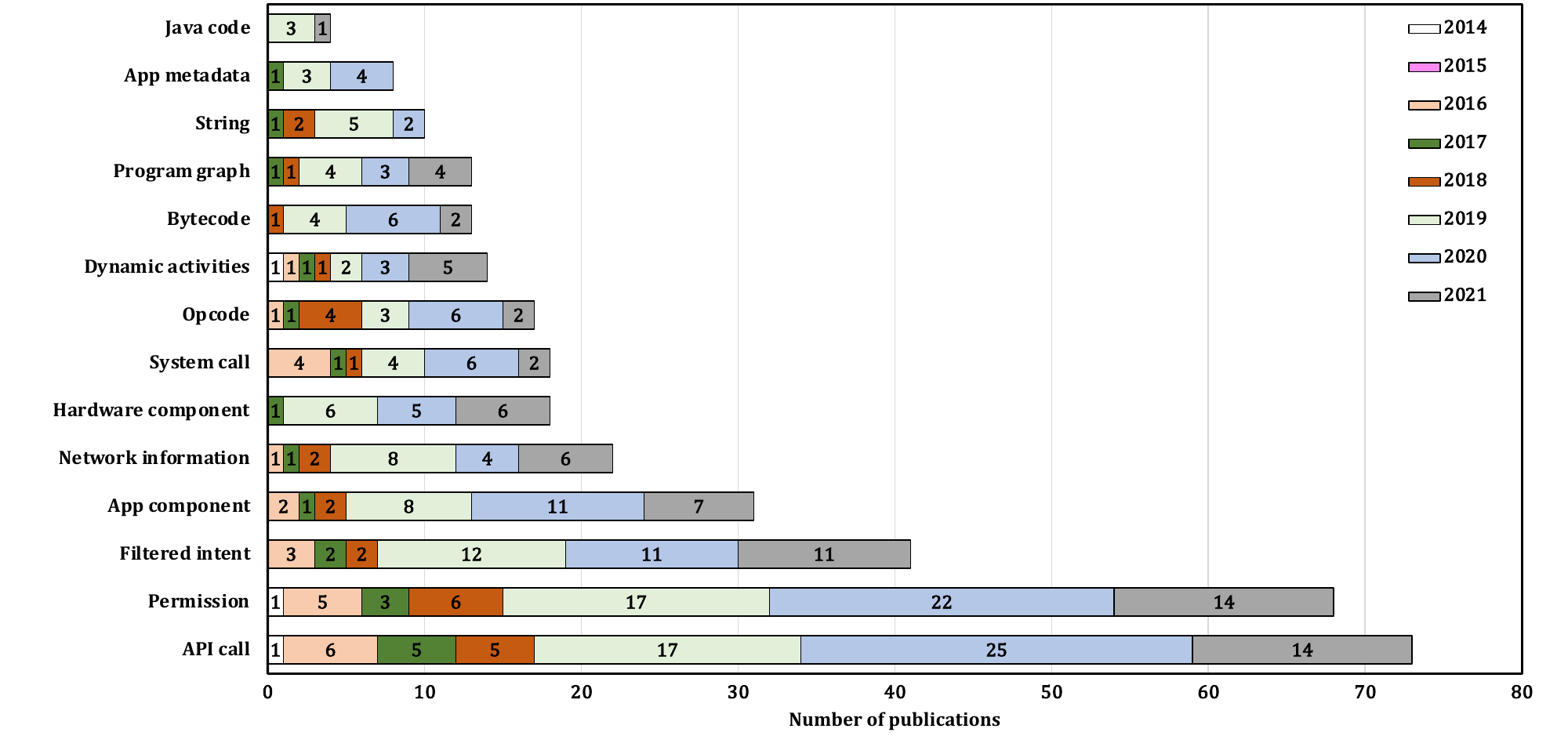}
  \caption{Summary of the primary studies by feature categories.}
  \label{fig:RQ2featurecategories}
\end{figure}

\subsubsection{Feature categories}
\label{sec:featurecategories}
As illustrated in Fig.~\ref{fig:RQ2featurecategories}, extracted features can be summarized into 13 categories, indicating the diversity of raw feature types.
Note that many studies may combine multiple types of features in order to accurately represent a malicious application.

As can be observed from Fig.~\ref{fig:RQ2featurecategories}, semantic features are the most common.
API calls (55.3\%) and permissions (51.5\%) have been the most frequently used feature types, accounting for well over half of primary studies. 
A possible explanation for this might be that API calls and permissions carry sufficient semantics and that the risk API calls and permissions usually result in dangerous or malicious behavior.
Other types of semantic information extracted from the decompiled code such as filtered intents and app components are also used by a large number of primary studies. 
There are also 13 primary studies (10\%) using program graphs like Control Flow Graph (CFG) and Data Flow Graph (DFG) to represent an application when analyzing Android malware.
Apart from the semantic information extracted from decompressed APK, we find eight recent studies leverage app metadata such as icons and app descriptions for the subsequent analysis. 

Although the aforementioned features are usually extracted via static analysis, we discover two distinct dynamic features.
18 primary studies employ Linux kernel system calls as extracted features to capture malicious behaviors.
Unlike API calls, Linux kernel system calls are not dependent on the Android OS version, making them more resilient to malware evasion strategies~\cite{hou2016deep4maldroid}.
Additionally, 14 primary studies examine characteristics associated with dynamic activities such as network access and memory dump.
These observations from Fig.~\ref{fig:RQ2featurecategories} corroborate those from Fig.~\ref{fig:RQ2programanalysis}, indicating that static analysis is the most frequently occurring approach for program analysis. 

Although high-level semantic features such as API calls remain the most commonly used, there is an increasing number of primary studies using raw code sequences to construct feature vectors.
Fig.~\ref{fig:RQ2featurecategories} indicates that the most frequently occurring raw code feature is raw opcode sequences from disassembled Android apps (with 22 studies).
The raw opcode sequences are fed into deep neural networks to learn high-level semantic feature representation automatically~\cite{qiu2020survey}.
Notice that four primary sources convert disassembled code to Java source code to construct feature vectors. 
On the other hand, we find that 13 primary studies fed the deep neural models with the raw classes.dex bytecode. 
For example, R2-D2~\cite{huang2019deep} converts bytecode into a color image by mapping the bytecode's hexadecimal value to the RGB color code.


\subsubsection{Feature encoding approaches}
\begin{figure}[t]
  \centering
  \includegraphics[width=1\textwidth]{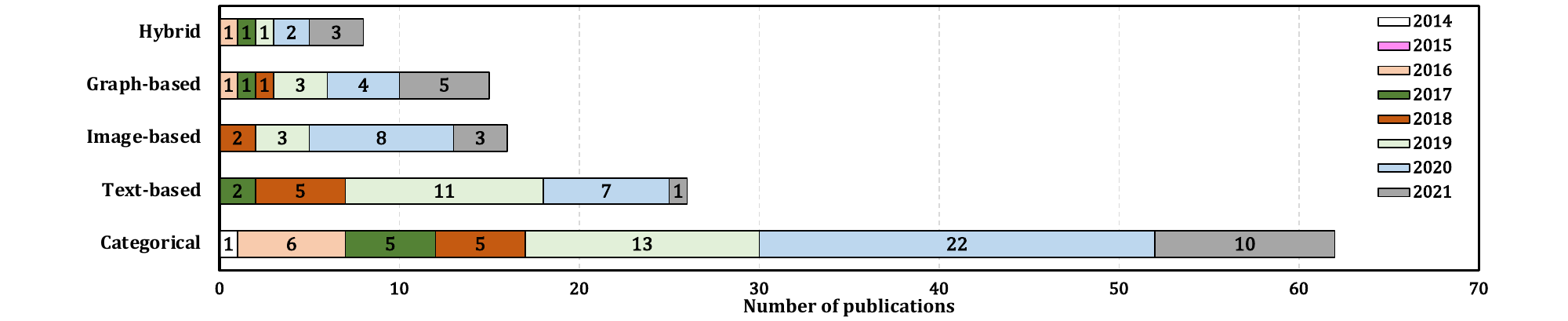}
  \caption{Summary of the primary studies by feature encoding approaches.}
  \label{fig:RQ2featureembedding}
\end{figure}

Fig.~\ref{fig:RQ2featureembedding} provides a summary of examined sources based on feature encoding approaches.
Following program analysis, the extracted information is further encoded into feature vectors and then fed into deep learning models. 
There are numerous ways to represent extracted features in primary studies, as extracted data from Android applications take on a variety of categories.
Thus, we classify feature encoding approaches into the following five categories:

\textbf{Categorical encoding.} 
Fig.~\ref{fig:RQ2featureembedding} indicates that categorical encoding approaches are most frequently occurring at 47\% of sources (62 primary studies).
This result appears to be consistent with Section~\ref{sec:featurecategories} which indicates that categorical semantic features like API calls and permissions are the most frequently used.
Typically, a numerical vector is constructed to indicate the presence of each categorical feature.
It is noteworthy that we discovered that 55 out of 62 primary studies adopt one-hot encoding to record the information of the presence of each possible feature value for applications.
For instance, DroidDetector~\cite{yuan2016droiddetector} considers a total of 192 features through hybrid analysis, and constructs a 192-dimensional vector for each app where each feature is assigned a value of 1 if it occurs in the app; otherwise, it is assigned a value of 0.
Besides, we find that seven sources assign each feature a discriminative integer and store the used features in a numerical vector.
Although categorical encoding is the most prevalent strategy because of its simplicity, it has two significant drawbacks: (1).high dimensional generation, (2).embedding in isolation between distinct patterns~\cite{li2018word}.

\textbf{Text-based encoding.} 
It is quite common to employ approaches from natural language processing to encoding sequential features.
Fig.~\ref{fig:RQ2featureembedding} indicates that 26 primary studies (20\%) attempt to utilize text-based feature encoding approaches. 
Numerous state-of-the-art text encoding approaches have been introduced to process sequential data.
In fact, one-hot encoding is the simplest method of text encoding but one of its disadvantages is high dimensional problem that we discussed before. 
In addition, some researchers employ discrete encoding approaches like Bag of Words (BOW), Term Frequency–Inverse Document Frequency (TF-IDF), and N-Gram~\cite{yen2019android, vinayakumar2017deep, xu2016hadm, qiu2019a3cm, ullah2019detection, ananya2020sysdroid, pierazzi2020data, tan2020end}. 
These methods, however, are still limited by data sparsity and high dimensionality issues~\cite{wang2020static}. 
Therefore, many primary studies further investigate the effectiveness of pre-trained word embedding models, such as Continuous Word2vec~\cite{karbab2018maldozer, zhang2019familial, wang2018deep, feng2019ac, chaulagain2020hybrid, zhu2020sadroid, bakhshinejad2019parallel} and GloVe~\cite{karbab2018maldozer}.

\textbf{Graph-based encoding.} 
We find that 15 primary studies (11\%) employ graph-based representation approaches.
Deep4MalDroid~\cite{hou2016deep4maldroid} obtains system calls through dynamic analysis tools to construct a weighted directed graph, and graph structure information including weights of each edge and in-degree and out-degree of each node is stored in vectors as inputs. 
Xu et al.~\cite{zhiwu2019android} encode CFG and DFG into adjacency metrics respectively and combine them into a single metric in embedding layers. 
In~\cite{pektacs2020deep}, the authors investigate several state-of-art graph embedding approaches to encode API call graphs, including DeepWalk~\cite{perozzi2014deepwalk}, Node2vec~\cite{grover2016node2vec}, HOPE~\cite{ou2016asymmetric}, etc. 

\textbf{Image-based encoding.} 
Image-based representation, employed in 16 primary studies (12\%), usually transforms extracted features into a grayscale or color image. 
The most common scenario is directly transforming bytecode into images.
For instance, IMCFN~\cite{vasan2020imcfn} reads an Android binary as a vector of 8-bit unsigned integers and then converts it into a two-dimensional array. 
Following that, the Android bytecode is visualized as a color image based on the RGB color map. 
Numerous research studies used similar approaches to encode Android bytecode~\cite{hsien2018r2, mercaldo2020deep, ren2020end, he2019malware, sun2019android, xiao2019image, bakour2020visdroid}.

\textbf{Hybrid encoding.} 
Combining distinct feature encoding approaches to process richer features is also common in collected research (6\%). 
Take Kim et al.~\cite{kim2018multimodal} as an example. 
The authors construct one-hot vectors to record the existence of categorical features like permission, string and app components. 
At the same time, in order to alleviate the impacts of obfuscation techniques, a similarity-based feature vector generation process is introduced to encode sequential features like opcode and API calls. 
In~\cite{xi2019deepintent,karunanayake2020multi, pei2020combining}, since these studies also consider icons or pictures of Android applications, both image embedding approaches and text embedding algorithms are used to encoding features.

~\\
\noindent \textbf{DISCUSSION.}   
According to our reviewed results, the majority of research constructs feature vectors by recording the existence of various categorical features of Android applications. 
Many studies create a look-up table to list all the potential features based on prior knowledge or feature selection approaches, and then build a fixed-size one-hot feature vector to represent each application~\cite{yuan2016droiddetector, yuan2014droid, wang2019effective, rathore2021robust, fereidooni2016anastasia, hou2016droiddelver, li2019adversarial, oak2019malware, feng2020performance, wang2016droiddeeplearner, feng2019mobidroid, martin2017evolving, hou2017deep, fan2020can, gong2020experiences, bai2020unsuccessful, bai2021comparative, wang2020sedroid, wu2020android, gronat2019maxnet, li2021robust}.
For instance, Wu et al.~\cite{wu2020android} identified 158 high-risk features to construct feature vectors (including 97 API calls and 61 permissions).
However, there are several issues to process features in this way. 
One of these is that it is pretty difficult to define a robust malicious feature list using either humans' experience or traditional feature selection approaches. 
The built feature lists can't encompass all potential malicious characteristics, resulting in poor performance in the practical application. 
Even when all features in the training data are used, concept drift caused by Android malware evolution is a serious problem that cannot be ignored~\cite{zhang2020enhancing}. 
Android malware continues to evolve with similar functionality but a completely different implementation, easily evading detection by Android malware defense models.
As a result, how to design effective and practical feature lists is a challenging issue. 

As shown in Fig.~\ref{fig:RQ2programanalysis}, static program analysis is the most common approach (73\%).
Furthermore, our results in Section~\ref{sec:featurecategories} show that the majority of reviewed studies extract static semantic features from disassembled files.
A significant drawback of this approach is its weak ability to handle obfuscation problems.
Obfuscation techniques (e.g., polymorphic code, encryption) transform malware binaries into self-compressed and uniquely structured binary files that are resistant to reverse-engineering approaches~\cite{gandotra2014malware, or2019dynamic}. 
Obfuscation techniques improve code protection for Android apps, but create significant barriers to malware analysis. 
For example, code reordering aims to modify the order of instructions in smali code but preserve the original run-time execution trace, thereby evading detection by malware defense tools~\cite{bacci2018detection}. 
By using a variety of obfuscation techniques, malware attackers can produce multiple variants of a single malicious sample, complicating malware defenses. 
Although some studies have shown that the proposed DL-based approaches are slightly affected by some simple obfuscation approaches~\cite{kim2018multimodal, xu2018deeprefiner, lee2019seqdroid, millar2020dandroid}, we cannot ignore the fact that the real-world obfuscation techniques constantly update and evolve against anti-malware approaches~\cite{qiu2020survey}.
Investigating obfuscated apps using deep learning techniques is a potential future research topic, and we outline some potential research trends: (1). using deep learning techniques to detect and analyze obfuscation approaches; (2). analyzing malware based on bytecode-level rather than capturing semantic features.

\begin{tcolorbox}
\textbf{RQ2.1 How are features processed for model training?} 
\begin{itemize}
    \item Static analysis is mostly used to obtain features, and static semantic features like API calls and permissions remain the most frequently utilized.
    \item The number of primary studies devoted to dynamic analysis is rising and many generally applicable methodologies/frameworks are proposed. 
    \item One-hot encoding and text encoding are mostly used to represent features. 
    \item 13 primary studies encode raw bytecode into feature vectors. 
\end{itemize} 
\end{tcolorbox}

\subsection{Deep Learning Techniques}
\label{sec:deeplearningtech}
Responding to RQ2.2, this section provides a detailed review of the primary studies according to deep learning techniques. 
To comprehend this section, readers are expected to be relatively familiar with deep learning.
For more details on the patterns described, readers are referred to the deep learning textbook by Goodfellow et al.~\cite{goodfellow2016deep}.

\begin{table}[t]
\scriptsize
  \centering
  \caption{A summary of learning paradigms}
    \begin{tabular}{lrrrr}
    \toprule
         & \multicolumn{1}{c}{\textbf{Supervised learning}} & \multicolumn{1}{c}{\textbf{Unsupervised \& supervised}} & \multicolumn{1}{c}{\textbf{Unsupervised learning}} & \multicolumn{1}{c}{\textbf{Reinforcement learning}} \\
    \midrule
    \textbf{Counts} & 108  & 20    & 1    & 3 \\
    \textbf{Percentage} & 81.8\% & 15.2\% & 0.8\% & 2.3\% \\
    \bottomrule
    \end{tabular}%
  \label{tab:RQ2learningparadigms}%
\end{table}%

\subsubsection{Learning paradigms}
Regarding the type of deep learning paradigms, Table~\ref{tab:RQ2learningparadigms} indicates that supervised learning-based Android malware defenses appear most frequently (81.8\%). 
It is worth noting that only one primary source employs unsupervised learning. 
Specifically, CADE~\cite{yang2021cade} proposes an unsupervised representation learning approach to combat concept drift for security applications. 
Twenty primary studies (15.2\%) develop Android malware defense approaches based on unsupervised \& supervised scenarios.  
Specifically, unsupervised DNN models such as Auto-Encoders are usually employed to initialize a neural network's weights. 
Then, the pre-trained model can be fine-tuned using labeled samples using a standard supervised back-propagation algorithm~\cite{yuan2014droid, su2016deep, fereidooni2016anastasia, hou2016deep4maldroid, wang2016droiddeeplearner, hou2016droiddelver, yuan2016droiddetector, xu2016hadm, hou2017deep, zhu2017deepflow, gharib2017dna, wang2019effective, Khoda2019mobile, lu2020android, su2020droiddeep, d2020malware, wang2020sedroid, zhu2021hybrid}. 
Besides, three primary studies rely on reinforcement learning to conduct the research~\cite{wan2017reinforcement, rathore2021robust, zhao2021structural}.
These observations indicate that supervised learning techniques that require sufficient labeled data occupy an absolutely dominant position in this research domain currently.

\begin{figure}[t]
  \centering
  \includegraphics[width=1\textwidth]{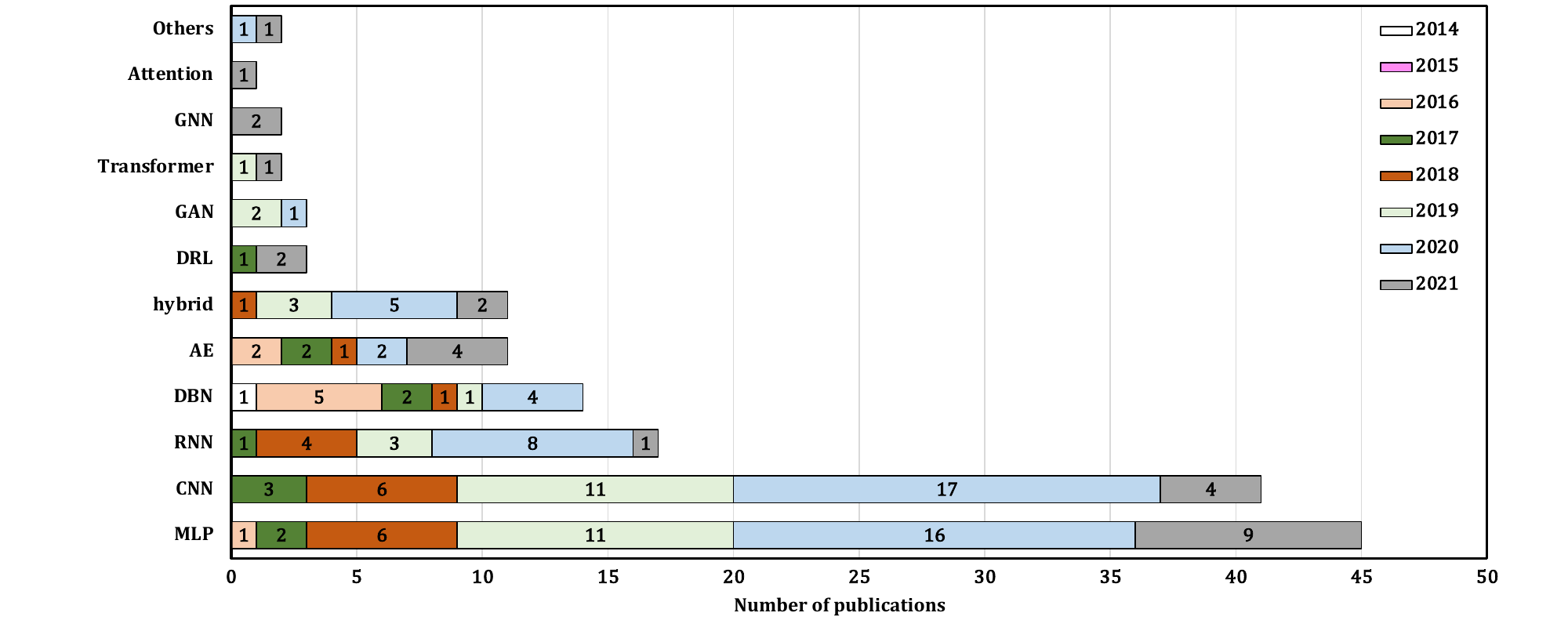}
  \caption{Summary of the primary studies by deep learning models.}
  \label{fig:RQ2DeepLearningModel}
\end{figure}

\subsubsection{Deep learning models}
The reviewed sources consider a variety of deep learning models.
We categorize these DL models according to their model architectures and summarize the primary studies regarding DL models in Fig.~\ref{fig:RQ2DeepLearningModel}.

\textbf{Multilayer Perceptrons (MLPs)}, also known as deep feedforward networks or feedforward neural networks, are among the simplest deep learning models for Android malware defenses. 
Building an MLP is straightforward and MLP can learn hierarchical feature representation of inputs.
MLPs are demonstrated to be universal approximators, capable of approximating any measurable function to any designed degree of accuracy~\cite{hornik1989multilayer}.
As a result, MLPs serve as the basis of many advanced deep learning models and are widely used in a variety of research areas~\cite{goodfellow2016deep}.
We thus observe from Fig.~\ref{fig:RQ2DeepLearningModel} that MLPs are the most frequently occurring DNN models, referring to 45 primary studies (34.1\%).

\textbf{Convolutional neural networks (CNNs, ConvNets)} improve traditional MLPs by introducing convolution and pooling (or subsampling) operations to learn high-level features from low patterns with higher efficiency and accuracy. 
It is remarkable that CNNs have also been among the most popular deep learning models in Android malware defenses, accounting for 41 sources (31.1\%) since 2017.
We observe that numerous sources employ CNNs to learn feature representations for opcode, bytecode, and API call sequences~\cite{mclaughlin2017deep, vasan2020imcfn, karbab2018maldozer, pektacs2020deep, hsien2018r2, nix2017classification, ye2019out, yuan2020byte, ding2020android, iadarola2021towards, he2019malware, bakour2020visdroid, li2020opcode, sun2019android, bakhshinejad2019parallel}. 
This observation is unsurprising given that CNNs can automatically learn useful context structural information of Android apps in comparison to MLPs.

\textbf{Recurrent Neural Networks (RNNs)} have emerged as a successful paradigm for modeling sequential data since RNNs incorporate hidden units that implicitly maintain the history of all past elements in the sequence~\cite{elman1990finding}.
As such, it becomes a powerful tool in Natural Language Processing (NLP) and speech processing~\cite{lecun2015deep}.
The programming pattern of Android applications is sequential and logical, and capturing sequential and semantic properties from the decompiled code is essential for improving malware defense models' performance.
Fig.~\ref{fig:RQ2DeepLearningModel} shows that 17 primary studies (12.9\%) use RNNs to defend against Android malware attacks. 
Except for standard RNNs, advanced variants such as Long Short Term Memory (LSTM) and Gated Recurrent Unit (GRU) are often deployed to overcome the vanishing gradient problem, thereby enhancing model performance~\cite{xiao2019android, vinayakumar2018detecting, jahromi2020enhanced, de2020visualizing, vinayakumar2017deep, yan2018lstm, chaulagain2020hybrid, zhou2020android}. 

\textbf{Autoencoders (AEs)} are unsupervised neural networks that learn the latent space of inputs in an unsupervised manner.
AEs have been successfully used in dimensionality reduction and information retrieval tasks~\cite{bengio2013representation, goodfellow2016deep}. 
The typical structure of AEs consists of two components: an encoder mapping inputs to a hidden representation, and a decoder mapping the hidden representation back.  
AE and its variants, such as Denoising Autoencoders (DAEs) and Variational Autoencoders (VAEs), have been widely applied to Android malware defenses, referring to 11 sources (8.3\%).

\textbf{Deep Belief Networks (DBNs)} belong to probabilistic generative models, and DBNs are hierarchically constructed by multiple layers of stochastic hidden variables \cite{hinton2009deep}.
DBNs learn multiple layers of features from unlabeled inputs in an unsupervised manner, and these features can then be used to optimize discrimination in a supervised manner to perform the classification tasks. 
Deep belief networks are one of the first non-traditional models to admit deep architecture training successfully, but they are currently rarely used compared with other advanced deep learning networks \cite{goodfellow2016deep}.
Fig.\ref{fig:RQ2DeepLearningModel} presents a consistent result that DBNs were also the first DNN network to build Android malware defense models, with the highest proportion between 2014 and 2016.

\textbf{Generative Adversarial Networks (GANs)} are composed of a generative model and a discriminative model.
The generator is trained to generate new samples in order to fool the discriminative model, while the discriminator tries to distinguish the generated samples from the real samples, liking in a game of cat and mouse to compete with each other. 
GANs were even described as the most interesting machine learning idea in the last ten years by AI pioneer Yann LeCun in 2016.
Nevertheless, only three primary studies render GAN-based Android malware defense architectures~\cite{millar2020dandroid, li2019adversarial, amin2019android}, showing that further explorations to the application of GANs in malware defenses are still needed.

\textbf{Graph Neural Networks (GNNs)} are designed by extending deep learning techniques to graph data. 
AI researchers have developed a variety of GNN architectures, such as Graph Convolutional Network (GCN)~\cite{kipf2016semi} and Graph Attention Network (GAT)~\cite{velivckovic2017graph}.
In Fig.~\ref{fig:RQ2DeepLearningModel}, we can observe that two recent studies propose GNNs-based heterogeneous graph representation learning approaches in Android malware defenses~\cite{fan2021heterogeneous, gao2021gdroid}. 

\textbf{Attention-based neural networks} are capable of learning the dependencies between inputs and target sequences, bringing a huge improvement in Machine Translation~\cite{weng2018attention}.
As such, one source~\cite{wu2020android} proposed an explainable attention-based Android malware detection model since attention mechanisms can provide information about the elements' relevance to their targets. 
Although there is currently only one primary study employing attention mechanisms, a variety of popular attention mechanisms have been proposed and demonstrated to perform well in NLP or CV, such as self-attention, soft/hard attention, local and glocal attention, co-attention, etc.
As a result, we suggest that future researchers make more efforts to apply attention-based models to more specific issues in Android malware defenses. 

\textbf{Deep Reinforcement Learning (DRL)} operates on a trial-and-error paradigm to teach an autonomous agent how to perform a task without human guidance.
DRL has demonstrated its significant performance in the fields of games, robotics, and self-driving cars~\cite{arulkumaran2017deep, franccois2018introduction}.
In Android malware defenses, we also find three DRL-based primary studies~\cite{bai2021comparative, rathore2021robust, zhao2021structural}. 
For example, Zhao et al.~\cite{zhao2021structural} and Rathore et al.~\cite{rathore2021robust} examine the effectiveness of reinforcement learning for evading adversarial attacks and relevant protection strategies. 

\textbf{Transformers} enable much more parallelization than CNNs and RNNs by leveraging the self-attention mechanism. 
This capability makes it possible to efficiently (pre-)train extremely large language models on GPUs. 
Bidirectional EncoderRepresentations from Transformers (BERT) is among the most widely used transformer-based pre-trained language models~\cite{devlin2018bert}.
We found one research study that employed BRET to model sequential features on highly imbalanced malware data. 

\textbf{Hybrid-based models} integrate several basic DNN blocks to formalize more robust and effective models.
As shown in Fig.~\ref{fig:RQ2DeepLearningModel}, hybrid-based models have been leveraged in 11 primary studies (8.3\%).
Numerous deep learning framework combinations have been considered, like AE and CNN~\cite{wang2019effective, karbab2021petadroid}, RNN and CNN~\cite{pektacs2020learning, pei2020combining}, MLP and LSTM~\cite{xu2018deeprefiner}, etc.

\textbf{Others.} Recent years have seen rapid advancements in deep learning, with new deep learning techniques being proposed constantly.
Two sources leverage deep learning models that fall outside of the aforementioned categories. 
Bai et al.~\cite{bai2020unsuccessful} perform Android malware family classification through siamese neural networks.
Ma et al.~\cite{ma2021deep} adopt deep residual learning to detect sensitive behaviors.

Fig.~\ref{fig:RQ2DeepLearningModel} demonstrates that multiple types of deep learning models have been adopted to defend against Android malware.
In the earlier years, unsupervised DNN models like AE and DBN drew the most research attention. 
However, beginning in 2017, popular supervised DNN architectures such as MLP, CNN, and RNN garnered increased attention. 
Another intriguing finding is that a variety of advanced approaches have emerged over the past three years, including GAN, GNN, attention, transformer, DRL, and hybrid models.
These observations further support that deep learning has garnered growing interest in the field of Android malware defenses.

\begin{table}[t]
  \centering
  \scriptsize
  \caption{A summary of explanation approaches used in DL-based Android malware defenses}
    \begin{tabular}{lcp{7em}p{24em}c}
    \toprule
    \multicolumn{1}{c}{\textbf{Tool/System.Ref}} & \textbf{Year} & \multicolumn{1}{c}{\textbf{DNN models}} & \multicolumn{1}{c}{\textbf{Explanation approach}} & \multicolumn{1}{c}{\textbf{Explanation type}} \\
    \midrule
    Zhu et al.~\cite{zhu2019transparent} & 2019  & CNN   & Global surrogate (training a simple CNN) & Global \\
    Pierazzi et al.~\cite{pierazzi2020data} & 2020  & MLP, DBN ,CNN & Mean Decreased Impurity & Global \\
    Fan et al.~\cite{fan2020can} & 2020  & MLP   & LIME, SHAP, LEMNE, Anchor, LORE & Local \\
    DENAS~\cite{chen2020denas} & 2020  & MLP   & DENAS (approximating the non-linear decision\newline{}boundary of DNNs using an iterative process) & Global \\
    Warnecke et al.~\cite{warnecke2020evaluating} & 2020  & MLP, CNN & Gradients and Integrated Gradients,  Layer-wise relevance propagation (LRP), LIME, SHAP, LEMNA & Local \\
    Feichtner et al.~\cite{feichtner2020understanding} & 2020  & CNN   & LIME  & Local \\
    XMal~\cite{wu2020android} & 2021  & Attention-based & Attention mechnism & Local \\
    Severi et al.~\cite{severi2021explanation} & 2021  & MLP   & SHAP  & Local \\
    Iadarola et al. ~\cite{iadarola2021towards} & 2021  & CNN   & Grad-CAM & Local \\
    CADE~\cite{yang2021cade} & 2021  & AE    & A distance-based approach (contrastive learning) & Local \\
    \bottomrule
    \end{tabular}%
  \label{tab:RQ2ModelExplain}%
\end{table}%

\subsubsection{Model explanation}
Deep learning approaches with a sophisticated architecture remain black-box models~\cite{molnar2020interpretable}.
Specifically, these constructed models cannot provide evidence to interpret why a given sample is identified as malicious.
The absence of sufficient transparency and trustworthiness in the proposed approaches is a significant obstacle to employing theoretical tools in practical malware analysis.
As a result, it is necessary to develop explanation approaches for malware defense models.
In collected studies, ten studies (7.5\%) exploited interpretable deep learning techniques in Android malware defenses.
Table~\ref{tab:RQ2ModelExplain} summarizes a comparative result towards interpretable tools. 
Interestingly, nine out of ten primary sources are proposed after 2019, indicating that explainable deep learning approaches for malware defenses are a current hot research topic.

With respect to the scope of interpretability, it is remarkable that most studies leverage local approaches (seven out of the 10 studies).
Global methods describe how features affect the prediction on average, whereas local methods aim to explain individual predictions~\cite{molnar2020interpretable}.
Regarding the explanation approaches used in sources,  state-of-the-art model-agnostic explanation approaches, including LIME~\cite{ribeiro2016should}, SHAP~\cite{lundberg2017unified}, Anchor~\cite{ribeiro2018anchors}, LORE~\cite{guidotti2018local}, and LEMNA~\cite{guo2018lemna}, are most frequently occurring (four primary sources~\cite{fan2020can, warnecke2020evaluating, feichtner2020understanding, severi2021explanation}).
These approaches treat the target classifier as a blackbox and approximate the decision boundary of any machine learning model by using a simple explainable model.
Note that two primary studies employ gradient-based explanation approaches (e.g., integrated gradients and Grad-CAM) to back-propagate gradients through the DNN in order to measure the sensitivity of each feature~\cite{warnecke2020evaluating, iadarola2021towards}.
Wu et al.~\cite{wu2020android} design an interpretable approach to classify Android malware by leveraging a customized attention mechanism.

~\\
\noindent \textbf{DISCUSSION.}
Our results indicate that supervised learning techniques occupy an absolutely dominant position in the current research.
However, this kind of learning involves data labeling, which is costly and requires domain-specific knowledge.
In Android malware defenses, Anti-Viruses (AVs) like VirusTotal are widely used to provide ground truth for experimental data.
We cannot ignore the following significant issues.
First, it may be convenient to use AVs to distinguish between malware and benign apps. 
However, AVs couldn't perform complex labeling tasks such as evolved malware labeling or malware behavior labeling, which still require substantial expertise.
Second, most of the AVs work on the signature, heuristic, and behavior-based detection engines~\cite{ye2017survey}. 
These approaches, however, are still time-consuming and human-dependent, and the more serious problem is that they cannot work well on future samples~\cite{ye2017survey}. 
Furthermore, one commercial AV may produce inconsistent results over time, or distinct AVs may produce different results, causing the ground-truth unreliable~\cite{botacin2021challenges}.
As a result, reliable data labeling for Android malware defenses may be a potential research topic.
We also encourage our fellow researchers to focus more on deep learning techniques requiring less human labor to annotate data, like active learning, semi-supervised learning, reinforcement learning, or unsupervised learning.

Although all studies we investigated in this review are related to deep learning, most studies employed neural networks with three to four layers. 
Training a very deep neural network on a small scale of data would cause severe overfitting~\cite{goodfellow2016deep}. 
Although shallow DNN networks have demonstrated promising performance in Android malware defenses, a deeper neural network is worth further exploring in this domain. 
On the other hand, pre-trained models trained on large amounts of data play an important role in CV and NLP domains, as they lower the barrier to applying these DNN models to real-world problems. 
With the explosive growth of the number of Android malware, it appears that it is not a good solution to train a DNN model from scratch every time a model is needed. 
A pre-trained DL model for Android malware can considerably bring convenience for the research in this domain. 

With the exponential growth of Android applications, the requirement for massive computational resources to achieve the desired performance is becoming an increasing issue in this domain.
In comparison to text and image, Android files are larger in size and have a more complex structure and feature processing by reverse-engineering tools is required, which is time-consuming.
Furthermore, current deep learning frameworks involve a considerable amount of computational resources to approach state-of-the-art performance~\cite{pouyanfar2018survey}.
As a result, improving the computational efficiency of DL-based Android malware defense approaches is a growing need. 

Interpretable or explainable deep learning-based Android malware defenses are also a future interesting topic~\cite{tantithamthavorn2020explainable,guidotti2018survey}. 
Recently, researchers have focused on conducting empirical studies to highlight the need of explainable AI/ML models for software engineering~\cite{jiarpakdee2020xai4se} and developing novel approaches for explainable AI/ML models for software engineering~\cite{jiarpakdee2020xai4se, peng2020defect, wattanakriengkrai2020linedp,pornprasit2021jitline, rajapaksha2021sqaplanner,krishna2020learning,peng2020defect}.
Although prior studies have attempted to employ local/global explainable approaches to provide explanations based on the Android characteristic-based features for each unknown sample~\cite{wu2020android}, there are still several issues requiring further exploration. 
First, current studies mainly focus on simple semantic characteristic-based features extracted from APK, so richer feature types and in-depth explanations of source code should be investigated further. 
Specifically, deep learning techniques have been widely utilized to analyze raw code but how to transform the unreadable code into semantic interpretation is an unsolved problem. 
On the other hand, an effective evaluation system for explainable Android malware defenses is currently unavailable, making it significantly more difficult for researchers to measure the quality of explanation results and compare alternative explanation approaches. 
Indeed, it is nearly impossible for experienced malware analysts to detect all malicious behaviors in malware samples without making a mistake.
Therefore, improving the reliability of explanations for malware samples is a potential challenge for future researchers. 

\begin{tcolorbox}
\textbf{RQ2.2: What deep learning architectures are used?} 
\begin{itemize}
    \item MLPs, CNNs, and RNNs are mostly used in Android malware defenses.
    \item Research is primarily focused on supervised learning tasks, especially binary classification tasks. 
    \item The applications of recent advanced DL techniques (e.g., GAN, attention and DRL) to Android malware defenses are still relatively preliminary.
    \item The interest in explainable DL-based malware defenses raises, and ten related studies have been published from 2019.   
\end{itemize} 
\end{tcolorbox}

\begin{table}[t]
  \centering
  \footnotesize
  \caption{A summary of the deployment of malware defense tools}
    \begin{tabular}{lrrr}
    \toprule
         & \multicolumn{1}{c}{\textbf{Off-device}} & \multicolumn{1}{c}{\textbf{On-device}} & \multicolumn{1}{c}{\textbf{Distributed}} \\
    \midrule
    \textbf{Number of papers} & 123  & 2    & 7 \\
    \textbf{Ratio} & 93.2\% & 1.5\% & 5.3\% \\
    \bottomrule
    \end{tabular}%
  \label{tab:RQ2deployment}%
\end{table}%

\subsection{Deployment of Analysis}
\label{sec:deploymentofanalysis}

Deployment approaches for malware defenses can be grouped into three categories: (i) off-device, (ii) on-device, and (iii) distributed, i.e., a combination of (i) and (ii)~\cite{faruki2014android}.

In 93.2\% of sources, the proposed tools are deployed off-device (see Table~\ref{tab:RQ2deployment}).
Specifically, most studies design an off-device approach and conduct experiments on personal computers or higher-performance GPU servers.
Automated malware defenses on a large volume of data require massive computational resources.
Thus, the majority of sources didn't consider deploying the obtained DL models on mobile devices for real usage. 

Conversely, only two studies propose on-device approaches~\cite{feng2019mobidroid,feng2020performance}.
On-device Android malware defenses provide analysis results through the mobile device itself, without the need to share or upload private data.
Currently, on-device Android malware defenses frameworks with deep learning techniques are usually implemented by transplanting the model trained on servers to smartphones. 
Feng et al.~\cite{feng2019mobidroid,feng2020performance} proposed two on-device Android malware detection systems, MobiDroid and MobiTive, which leveraged deep learning techniques to provide real-time detection on the user's mobile device. 
Specifically, they first maintained an effective Android malware detection model on the server side before migrating the pre-trained model to TensorFlow-lite \footnote{https://www.tensorflow.org/lite/} model. 
They demonstrated that the proposed approach could provide a reliable and fast reactive detection service on mobile devices. 

We also found seven distributed approaches (5.3\%)~\cite{wan2017reinforcement, hsien2018r2, alshahrani2018ddefender, ullah2019detection, ye2019out, tan2020end, gong2020experiences}.
Distributed malware defense that performs on-the-fly analysis and/or detection on the mobile device while performing detailed and computationally expensive analysis on a remote server \cite{faruki2014android}.
A good example is R2-D2 proposed by Hsien et al. \cite{hsien2018r2} in which Android users scan a suspicious app on their own mobile device, and if the app is previously unrecognized, the app's classes.dex is transformed into an RGB image that needs to be uploaded to the server side. 
In their back-end server, the image will be fed into a CNN network and, once identified, the results will be sent to the users' phone. 
One of the major drawbacks is that distributed approaches reveal significant private data to the cloud as the uploading process via the Internet and the analysing server itself may not be secure. 
In the machine learning domain, G{\^a}lvez et al.~\cite{galvez2020less} utilize semi-supervised ensemble learning to implement a privacy-respect malware detector. 
However, privacy protection for distributed DL-based approaches still needs to be thoroughly studied.

In commenting on these findings, we would encourage authors to propose more on-device and distributed malware defense approaches.
Furthermore, we would encourage future authors to investigate possible solutions to real-world issues such as privacy protection and computation resource limitations.

~\\
\noindent \textbf{DISCUSSION.}
Most of the surveyed studies proposed off-device Android malware defense models. 
They firstly collected a number of malware samples and performed model training and model evaluation on personal computers or GPU servers, which is a fairly common operation in the domain of deep learning applications. 
However, as malware techniques evolve and update, the problem of model aging is inevitable, resulting in significant performance degradation over time.
When a new Android malware family is reported, it is relatively difficult for these obsolete models to update it in a responsive time. 
In addition, off-device analysis can't provide timely protection for mobile users. 
Distributed or on-device Android malware is one of the potential solutions, but the number of related research studies is relatively small at the moment. 
Furthermore, the prior approaches suffer from several critical flaws requiring further investigations.

The existing distributed/on-device frameworks for Android malware defenses are quite simple and limited.
Firstly, suspicious Android applications must be uploaded to the server-side, which requires a heavy communication overhead. 
It is a potential option to assign parts of the computational tasks to smartphones.
But there is a challenge to seeking the trade-off between the detection performance and the real-time demand.
On the other hand, the communication process between clients and servers via the Internet may not be secure enough. 
It is not a difficult job for attackers to modify uploading data or steal private data. 
As a result, taking privacy protection into account is necessary for future work. 
Without a doubt, with the rapid development of deep learning techniques and smartphones, there will be new available DNN architectures supporting working effectively on mobile devices. 
Thus, we hope future researchers can propose more practical on-device approaches.

\begin{tcolorbox}
\textbf{RQ2.3 How are DL-based Android malware defenses approaches deployed in practice?} 
\begin{itemize}
    \item Most studies propose the Android malware defense models based off-device.
    \item There are seven papers focusing on distributed malware defenses.
    \item Only two papers propose on-device models that can perform a whole malware defense process on the mobile device.
\end{itemize} 
\end{tcolorbox}

\subsection{Performance Evaluation}
\label{sec:PerformanceEvaluation}
Assessing the performance of the proposed approach is an important process.
As a response for RQ2.4, we analyzed the evaluation approaches utilized in the surveyed studies. 

\begin{figure}[t]
  \centering
  \includegraphics[width=1\textwidth]{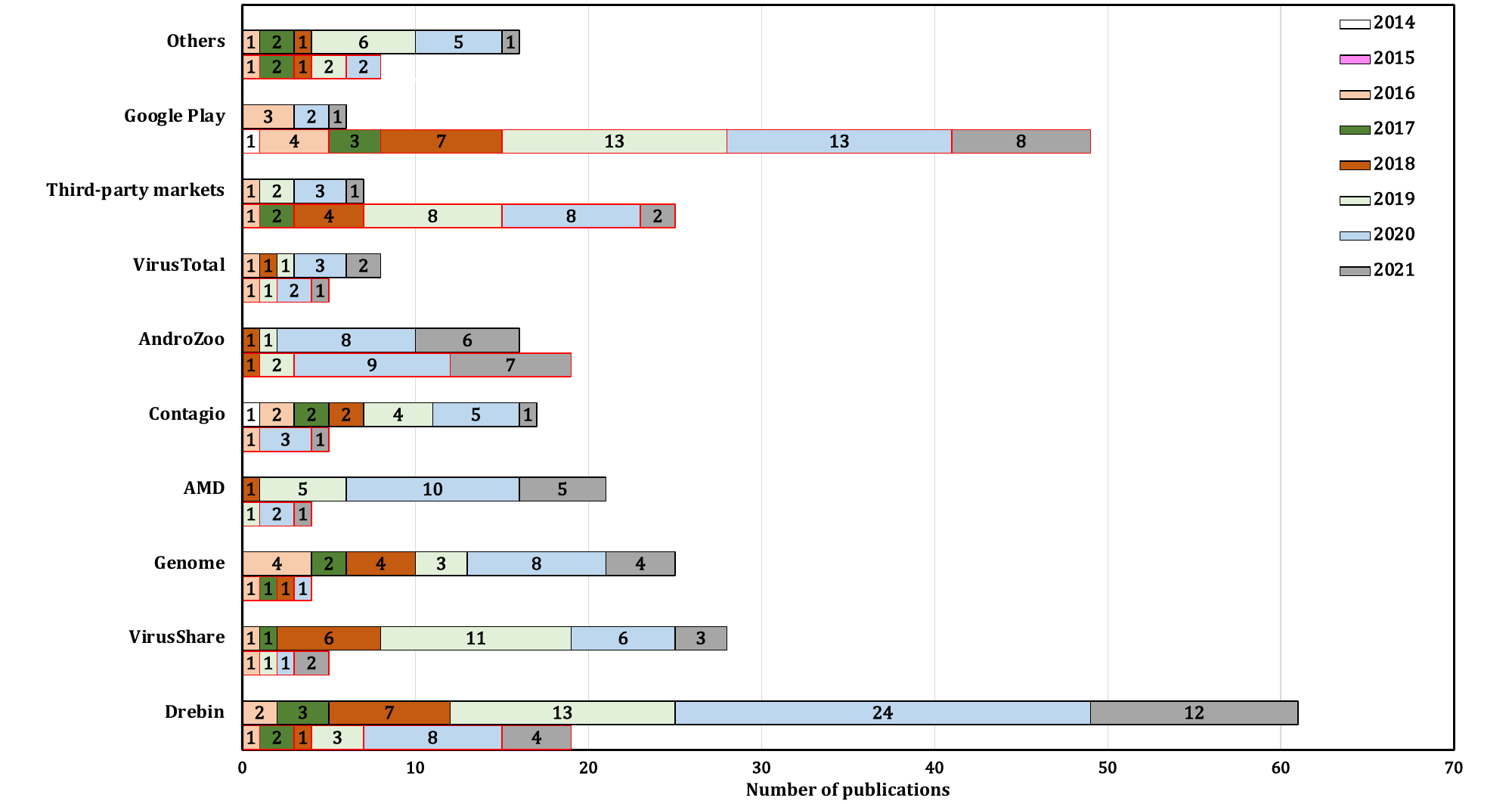}
  \caption{Summary of the primary studies by evaluation datasets (Red frame for goodware and black frame for malware). }
  \label{fig:RQ3EvaluationDatasets}
\end{figure}

\subsubsection{Dataset} 
\label{sec:dataset}

First, we examined the experimental datasets used in collected studies.
Fig.~\ref{fig:RQ3EvaluationDatasets} indicates that the authors can collect malware and goodware samples from a variety of sources. 
With respect to goodware samples, the official Google Play Store is the most frequently used one (37.1\%).
Note that third-party markets such as Anzhi, HUAWEI app store, and APKPure are also popular sources for real-world Android samples.
Conversely, public research datasets such as Drebin~\cite{arp2014drebin}, AMD~\cite{wei2017deep}, and Genome~\cite{zhou2012dissecting} are more popular to gather malicious applications.
It is remarkable that 61 primary studies collect malware samples from Drebin (46.2\%).
One potential disadvantage is that these datasets are not maintained or updated after being released, causing collected samples to be outdated.
Taking Drebin as an example, the dataset includes 123,453 benign samples and 5,560 malware samples from 2011 to 2014. 
Although these datasets are widely used, it appears that the evaluation results may not reflect the real detection capability of recent malware samples.
To overcome such a limitation, Fig.~\ref{fig:RQ3EvaluationDatasets} shows that the authors show an increasing interest in online repositories like AndroZoo~\cite{AndroZoo2020} and VirusShare~\cite{virusshare2020} to collect recent malicious samples.

\begin{figure}[t]
  \centering
  \subfigure[The distribution of the number of evaluated applications]{\includegraphics[width=0.36\textwidth]{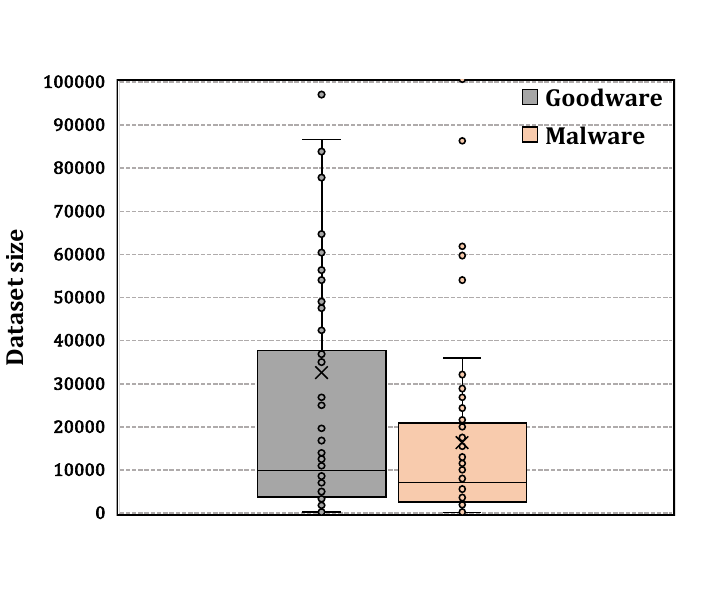}}
  \subfigure[The distribution of the ratio of goodware to malware]{\includegraphics[width=0.56\textwidth]{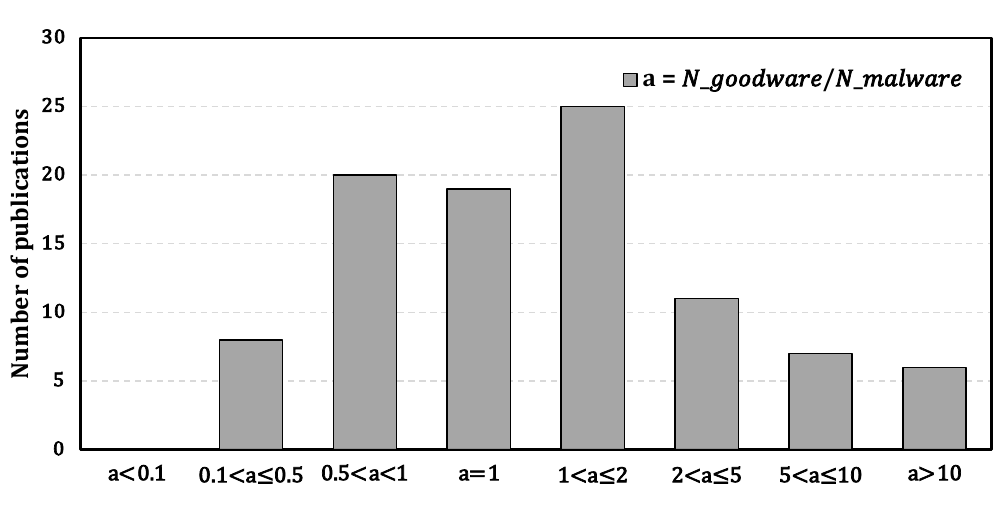}}
  \caption{Summary of the size of evaluation data}
  \label{fig:RQ3datasize}
\end{figure}

Second, for the scale of evaluation data, Fig.~\ref{fig:RQ3datasize}(a) represents the distribution of the number of evaluation data, while Fig.~\ref{fig:RQ3datasize}(b) displays the distribution of the ratio of the size of benign samples to malware samples. 
The median number of goodware samples used in performance evaluation is 9945, while the median number of malware samples used in performance evaluation is 7149 (see Fig.~\ref{fig:RQ3datasize}(a)).
Fig.~\ref{fig:RQ3datasize}(b) shows that the goodware:malware rate of 19 primary studies is set to 1:1, constructing a balanced dataset. 
Pendlebury et al.~\cite{pendlebury2019tesseract} described that the number of goodware in the real world is much greater than the number of malware, and Android malware accounts for between 6\% and 18.6\% of all apps.
However, it appears that only seven sources (5<a<=10) and six sources (a>10) appear to adhere to a realistic setting for the ratio of goodware to malware. 
There are even 28 primary studies constructing the evaluation dataset with more malware samples. 
Pendlebury et al.~\cite{pendlebury2019tesseract} confirmed that unrealistic assumptions about the ratio of goodware to malware cause biased performance. 
As such, we would encourage authors to construct evaluation data in appropriate and reliable settings.

\begin{figure}[t]
  \centering
  \includegraphics[width=1\textwidth]{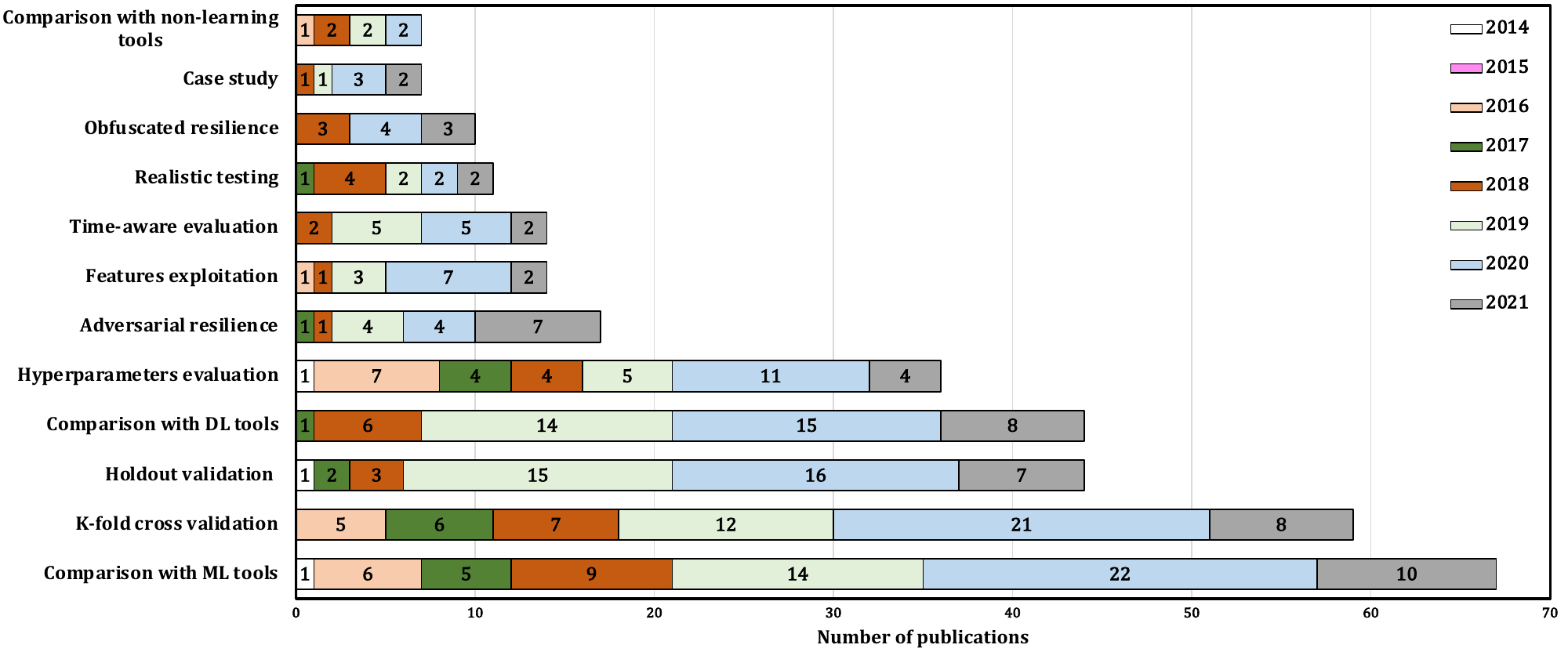}
  \caption{Summary of the primary studies by evaluation approaches.}
  \label{fig:RQ3EvaluationApproaches}
\end{figure}

\subsubsection{Evaluation approaches} 
Fig.~\ref{fig:RQ3EvaluationApproaches} summarizes the evaluation approaches used across the reviewed studies. 
We can observe here that 59 primary studies (44.6\%) employ k-fold cross-validation and 44 primary studies (33.3\%) utilize holdout validation.
Due to malware evolution, 14 primary studies (10.6\%) split evaluation data based on timestamps in order to perform time-aware experiments. 
To compare performance, we can see that 67 sources (50.7\%) compare the performance with traditional machine learning approaches like Drebin~\cite{arp2014drebin} and MaMaDroid~\cite{mariconti2016mamadroid}, while 44 sources (33.3\%) compare the performance with other deep learning-based approaches. 
It is interesting to observe that seven primary studies (5.3\%) compare the proposed approaches with non-learning tools such as signature-based anti-virus scanners. 
It is worth noting that 36 sources (27.2\%) evaluate the influence of hyperparameters in deep learning models (e.g., learning rate and hidden layer size).
Only seven primary studies conduct case studies to conduct an in-depth manual analysis of specific results.
To prove the robustness of the proposed approaches, some studies also examine their resilience against obfuscation (7.5\%) and adversarial attacks (12.8\%). 
To demonstrate the reliability of the proposed approaches, 14 primary studies (10.6\%) list the top significant features to measure the contribution of different features for the predictions.
Additionally, we discover that 11 sources (8.3\%) perform evaluation tests in real-world scenarios to further prove the performance.

\begin{figure}[t]
  \centering
  \includegraphics[width=0.55\textwidth]{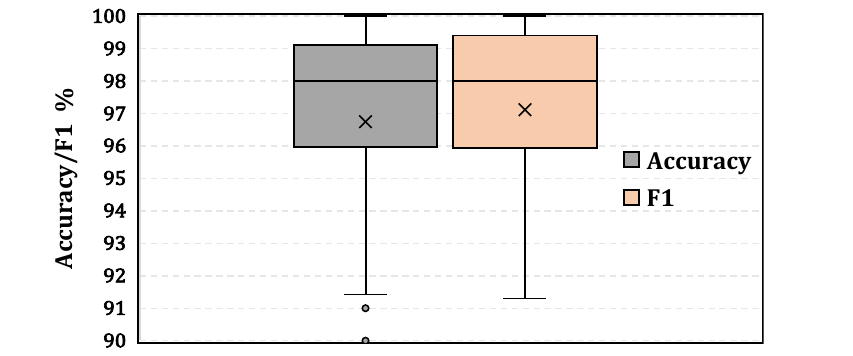}
  \caption{The distribution of the performance metric values.}
  \label{fig:RQ3EvaluationMetrics}
\end{figure}

\subsubsection{Evaluation metrics}
The previous results indicate that most studies focus on a classification problem.
As such, evaluation metrics from traditional classification problems are directly utilized to measure the performance of proposed malware defense models in the majority of collected studies. 
These standard classification evaluation metrics (e.g., accuracy, precision, recall, F1-score, True Positive Rate, False Positive rate, Receiver Operating Characteristic (ROC) curve) are discussed in \cite{qiu2020survey,liu2020review} in detail.
Computational efficiency is another common measure criterion in review papers.
Specifically, time costs like feature processing time and hardware resource consumption like memory usages are typically considered.

To ascertain the effectiveness of these deep learning-based approaches, we also looked at the specific values of the evaluation metrics.
Because accuracy and F1 score are the most frequently occurring (71.2\% sources use accuracy and 43.1\% sources use F1 score), we record the highest accuracy and F1 score values presented in reviewed sources.
Fig.~\ref{fig:RQ3EvaluationMetrics} presents their distribution. 
It is remarkable that the median number of accuracy/F1 is 98\%. 
Moreover, the accuracy/F1 values of 25\% of sources are greater than 99\%.
Numerous research studies~\cite{pendlebury2019tesseract, arp2020and} have pointed out that most studies present overly-optimistic results caused by a series of biased experimental settings (e.g., data imbalance, outdated data).
However, this issue still needs to be thoroughly studied in the future. 
For more information about potential risks/biases that undermine the evaluation performance, readers are referred to the empirical study by Arp et al.~\cite{arp2020and}.

\begin{table}[t]
  \caption{ List of publicly available tools}
  \scriptsize
  \centering
  \label{tab:RQ3Toolavailability}
    \begin{tabular}{lcp{5.7em}p{5.75em}lcp{5.7em}p{5.3em}}
    \toprule
    \multicolumn{1}{c}{\textbf{Tool}} & \textbf{Year} & \textbf{DNN models} & \textbf{Open-source tool-support} & \multicolumn{1}{c}{\textbf{Tool}} & \textbf{Year} & \textbf{DNN models} & \multicolumn{1}{p{6.5em}}{\textbf{Open-source tool-support}} \\
    \midrule
    McLaughlin et al.~\cite{mclaughlin2017deep} & 2017 & CNN  & \checkmark & Feichtner et al.~\cite{feichtner2020understanding} & 2020 & CNN  & \multicolumn{1}{p{6.5em}}{\checkmark} \\
    Kim et al.~\cite{kim2018multimodal} & 2018 & MLP  & \checkmark & Warnecke et al.~\cite{warnecke2020evaluating} & 2020 & MLP, CNN & \multicolumn{1}{p{6.5em}}{\checkmark} \\
    R2-D2~\cite{hsien2018r2} & 2018 & CNN  & \multicolumn{1}{l}{} & APICHECKER~\cite{gong2020experiences} & 2020 & MLP  &  \\
    Vinayakumar et al.~\cite{vinayakumar2018detecting} & 2018 & LSTM & \checkmark & DENAS~\cite{chen2020denas} & 2020 & MLP  & \multicolumn{1}{p{6.5em}}{\checkmark} \\
    TESSERACT~\cite{pendlebury2019tesseract} & 2019 & MLP  & \multicolumn{1}{l}{} & XMal~\cite{wu2020android} & 2021 & Attention & \multicolumn{1}{p{6.5em}}{\checkmark} \\
    Yen et al.~\cite{yen2019android} & 2019 & CNN  & \checkmark & Li et al.~\cite{li2021framework} & 2021 & AE, MLP & \multicolumn{1}{p{6.5em}}{\checkmark} \\
    Abderrahmane  et al.~\cite{abderrahmane2019android} & 2019 & CNN  & \checkmark & RAMDA~\cite{li2021robust} & 2021 & Hybrid & \multicolumn{1}{p{6.5em}}{\checkmark} \\
    DeepIntent~\cite{xi2019deepintent} & 2019 & Hybrid & \checkmark & Dr.Droid~\cite{fan2021heterogeneous} & 2021 & transformers, GNN & \multicolumn{1}{p{6.5em}}{\checkmark} \\
    ANDRE~\cite{zhang2019familial} & 2019 & MLP  & \checkmark & PetaDroid~\cite{karbab2021petadroid} & 2021 & Hybrid & \multicolumn{1}{p{6.5em}}{\checkmark} \\
    Karunanayake et al.~\cite{karunanayake2020multi} & 2020 & CNN  & \checkmark & DexRay~\cite{daoudi2021dexray} & 2021 & CNN  & \multicolumn{1}{p{6.5em}}{\checkmark} \\
    Li et al.~\cite{li2020adversarial} & 2020 & MLP  & \checkmark & Li et al.~\cite{li2021can} & 2021 & MLP, CNN, RNN & \multicolumn{1}{p{6.5em}}{\checkmark} \\
    BIHAD~\cite{pei2020combining} & 2020 & Hybrid & \checkmark & Severi et al.~\cite{severi2021explanation} & 2021 & MLP  & \multicolumn{1}{p{6.5em}}{\checkmark} \\
    API-GRATH~\cite{zhang2020enhancing} & 2020 & MLP  & \checkmark & Iadarola et al. ~\cite{iadarola2021towards} & 2021 & CNN  & \multicolumn{1}{p{6.5em}}{\checkmark} \\
    Shar et al.~\cite{shar2020experimental} & 2020 & MLP,CNN,RNN & \checkmark & CADE~\cite{yang2021cade} & 2021 & AE   & \multicolumn{1}{p{6.5em}}{\checkmark} \\
    Chaulagain et al.~\cite{chaulagain2020hybrid} & 2020 & LSTM & \checkmark & HRAT~\cite{zhao2021structural} & 2021 & DRL  &  \\
    \bottomrule
    \end{tabular}%
\end{table}%

\subsubsection{Availability}
The availability of proposed works in collected studies is investigated, which helps future researchers measure and certify the proposed tools. 
Table~\ref{tab:RQ3Toolavailability} presents a summary of the publicly available tools. 
In reviewed sources, there are only 30 primary studies providing publicly available tools, with just 22.7\%.
Among the publicly available approaches, 26 approaches are open-sourced.
It is noticeable that 26 sources out of the 30 are from 2019 to 2021, indicating that the research community in DL-based Android malware defenses is showing an increasing interest in sharing their research efforts. 

~\\
\noindent \textbf{DISCUSSION.}
The first problem is that evaluation results may not reflect the real performance of the model. 
For example, it is quite common to combine malware datasets released a long time ago such as Drebin and Contagio, with the latest benign samples from official Android markets to build new experimental trained data \cite{ding2020android, ananya2020sysdroid, millar2020dandroid, de2020visualizing}. 
These malware datasets have not been updated in real time to include the latest malware samples, and thus contain outdated malware samples. 
On the other hand, as defense strategies evolve, some malicious technologies used in public data sets are likely to be abandoned. 
The model trained using these types of data is rather weak to deal with the most recent malicious apps. 

The data distribution is highly imbalanced, posing a serial of challenges when performing malware analysis. 
First, the amount of benign samples is much larger than that of malware samples in the real-world scenario but this fact is usually overlooked by prior researchers~\cite{pendlebury2019tesseract}.
Similarly, the number of samples differs greatly among various malware families or malware categories. 
For Drebin data set as an example, though its 5560 malware samples can be classified into 179 distinct families, there are only 33 families that include more than 15 examples. 
Thus, malware families with small scale are easy to be misclassified into the training data set's dominant family categories.
Bai et al.~\cite{bai2020unsuccessful} demonstrated that although many proposed approaches have been proven good performance with high accuracy, these models suffered from poor performance to predict small families, even with the help of down-sampling methods. 
As a result, we suggest that considering data imbalance when analyzing malware may be critical for developing a more practical malware defense approach.

Another significant issue is limited reproducibility. 
Our reviewed results reveal that only a small number of primary studies share their source code, which makes future researchers' validation of assessment results more difficult.
Except for the source code, most studies collect evaluation data from a variety of sources but more details for the collected data are missing.
Daoudi et al.~\cite{daoudi2021lessons} attempted to reproduce five ML-based Android malware detectors, but only one was successful.
Thus, providing a transparent framework to manage performance evaluation is required for future researchers. 

\begin{tcolorbox}
\textbf{RQ2.4 How are DL-based Android malware defenses approaches evaluated?} 
\begin{itemize}
    \item Evaluation data is collected from a variety of resources. 
    \item Classic classification evaluation metrics are mostly used to measure the performance.
    \item Almost half of the sources reported a 98\% accuracy.
    \item Thirty works have made their contributions publicly available.
\end{itemize} 
\end{tcolorbox}

\begin{table}[t]
  \centering
  \scriptsize
  \caption{Top 10 cited papers based on citations per year}
  \label{tab:RQ3topcitedpapers}
    \begin{tabular}{lcccp{6em}p{23em}}
    \toprule
    \multicolumn{1}{c}{\textbf{Tool }} & \textbf{Year} & \textbf{Cites} & \textbf{Cit./Tear} & \multicolumn{1}{c}{\textbf{DNN models}} & \multicolumn{1}{c}{\textbf{Objectives}} \\
    \midrule
    Grosse et al.~\cite{grosse2017adversarial} & 2017  & 393   & 78.6  & MLP   & Adversarial Learning Attacks and Protections \\
    DL-Droid~\cite{alzaylaee2020dl} & 2020  & 130   & 65.0  & MLP   & Malware Detection \\
    McLaughlin et al.~\cite{mclaughlin2017deep} & 2017  & 313   & 62.6  & CNN   & Malware Detection \\
    DroidDetector~\cite{yuan2016droiddetector} & 2016  & 341   & 56.8  & DBN   & Malware Detection \\
    Kim et al.~\cite{kim2018multimodal} & 2018  & 219   & 54.8  & MLP   & \multicolumn{1}{l}{Malware Detection} \\
    IMCFN~\cite{vasan2020imcfn} & 2020  & 107   & 53.5  & CNN   & Malware Family Attribution \\
    Droid-Sec~\cite{yuan2014droid} & 2014  & 414   & 51.8  & DBN   & \multicolumn{1}{l}{Malware Detection} \\
    MalDozer~\cite{karbab2018maldozer} & 2018  & 207   & 51.8  & CNN   & Malware Detection, Malware Family Attribution \\
    Wang et al.~\cite{wang2019effective} & 2019  & 154   & 51.3  & hybrid & \multicolumn{1}{l}{Malware Detection} \\
    TESSERACT~\cite{pendlebury2019tesseract} & 2019  & 149   & 49.7  & MLP   & Malware Evolution Detection and Defense \\
    \bottomrule
    \end{tabular}%
\end{table}%

\begin{table}[t]
  \centering
  \scriptsize
  \caption{A summary of recent papers published in top venues}
    \begin{tabular}{lcp{7em}p{8em}p{22.165em}}
    \toprule
    \multicolumn{1}{c}{\textbf{Tool }} & \textbf{Year} & \multicolumn{1}{c}{\textbf{Venues}} & \multicolumn{1}{c}{\textbf{DNN models}} & \multicolumn{1}{c}{\textbf{Objectives}} \\
    \midrule
    TESSERACT~\cite{pendlebury2019tesseract} & 2019  & USENIX Security & MLP   & Malware Evolution Detection and Defense \\
    AiDroid~\cite{ye2019out} & 2019  & IJCAI & CNN   & Malware Detection \\
    DeepIntent~\cite{xi2019deepintent} & 2019  & CCS   & CNN, RNN, attention, AE & Malicious Behavior Analysis \\
    API-GRATH~\cite{zhang2020enhancing} & 2020  & CCS   & MLP   & Malware Evolution Detection and Defense, Malware Detection \\
    Karunanayake et al.~\cite{karunanayake2020multi} & 2020  & TMC   & CNN   & Repackaged/Fake App Detection \\
    DENAS~\cite{chen2020denas} & 2020  & FSE   & MLP   & Malware Detection \\
    Bai et al.~\cite{bai2020unsuccessful} & 2020  & ICSE  & Siamese network & Malware Family Attribution \\
    Severi et al.~\cite{severi2021explanation} & 2021  & USENIX Security  & MLP   & Adversarial Learning Attacks and Protections \\
    CADE~\cite{yang2021cade} & 2021  & USENIX Security  & AE    & Malware Evolution Detection and Defense \\
    XMal~\cite{wu2020android} & 2021  & TOSEM & Attention & Malware Detection \\
    Ficco et al.~\cite{ficco2021malware} & 2021  & TC    & MLP   & Malware Detection \\
    Dr.Droid~\cite{fan2021heterogeneous} & 2021  & SIGKDD & Transformers, GNN & Malware Evolution Detection and Defense, Malware Detection \\
    HDNFDroid~\cite{zhu2021hybrid} & 2021  & TKDE  & AE    & Malware Detection \\
    RAMDA~\cite{li2021robust} & 2021  & WWW   & MLP,  AE & Adversarial Learning Attacks and Protections \\
    HRAT~\cite{zhao2021structural} & 2021  & CCS   & MLP   & Adversarial Learning Attacks and Protections \\
    \bottomrule
    \end{tabular}%
  \label{tab:RQ3highrecentpaper}%
\end{table}%

\subsection{Trend Analysis}\label{sec:trendanalysis}
Advanced deep learning techniques have been widely applied in Android malware defenses since 2014. 
Fig.~\ref{fig:ResultAnalysis} reveals there is a growing interest in this emerging research topic, far from reaching its peak. 
In order to determine general research trends, we conduct a statistic analysis of the sources and present a detailed analysis for some specific popular topics as a response to RQ3.

\subsubsection{Statistic analysis}
In order to locate current research concerns, we first sorted collected studies according to the number of citations per year, and the counts are based on the citation counts of Google Scholar before Dec 2021. 
Table~\ref{tab:RQ3topcitedpapers} mentions the top 10 primary publications.
It is apparent from this table that integrating deep learning techniques into malware detection is still the main focus (seven out of the ten sources). 
But specifically, these research studies employ various deep learning models (e.g., MLP, CNN, DBN.) to detect malicious applications.
Note that Grosse et al.~\cite{grosse2017adversarial} has the largest number of citations, and one reason for the high citation counts is that this work investigates the viability of adversarial attack techniques against neural networks in this field, which sets the foundations for future work. 
Another work~\cite{pendlebury2019tesseract} published in 2019 that examines malware evolution is also a highly cited paper. 
These facts demonstrate that the robustness of DL-based detection models is attracting increased research attention. 

Although sorting these primary studies based on citations can help us identify high-impacted work, it is not applicable to recent work due to the time delay. 
We, therefore, list all recent publications (from 2019 to 2021) presented in top venues that have the highest quality ranking (A*) in the CORE ranking system, as shown in Table~\ref{tab:RQ3highrecentpaper}, which can help us seek the current research focus of top researchers. 
It is worth noting that top venues, especially in the security domain (CCS and USENIX Security), publish an increasing number of relevant studies.
Although most listed papers in this table still focus on malware detection, they make deeper analysis on more specific issues in Android malware detection, such as evaluation metrics, data imbalance problems, interpretability, model aging, etc. 
It is also noteworthy that "Adversarial Learning Attacks and Protections" and "Malware Evolution Detection and Defense" are frequently occurring in the top venues recently. 
These two issues are closely related to the practicability and effectiveness of the proposed architectures, which belong to key areas for future research to improve the state of the art in Android malware defenses. 

Table \ref{tab:RQ3topcitedpapers} and \ref{tab:RQ3highrecentpaper} further summarize the deep learning architectures used in these listed studies. Fig.~\ref{fig:RQ2DeepLearningModel} displays that CNN, RNN, and MLP have been widely used in Android malware defenses in recent years, MLP appears most frequently among listed important works. 
This is not a surprising outcome. MLP is the simplest but quintessential deep neural network, and thus researchers would tend to conduct experiments based on MLP networks first when they have a new research idea.
Compared with the sources listed in Table~\ref{tab:RQ3topcitedpapers}, recent studies adopt more advanced deep learning models such as attention-based networks and GNN.
These observations also illustrate that the application of advanced deep learning techniques to the Android malware defenses domain is actually in a preliminary stage.
In recent years, deep learning has brought impressive progress in many domains and many modern DL approaches are up-to-come, such as deep active learning, reinforcement learning, transfer learning, controllable generative models, etc. 
However, the application of deep learning technologies in the field of mobile security is far behind the development of deep learning itself. 
Research toward DL-based Android malware defenses is active currently, and hence we believe more advancements are expected in the near future. 

\begin{table}[t]
  \centering
  \footnotesize
  \caption{A summary of adversarial attacks and defenses}
    \begin{tabular}{p{4cm}p{2cm}p{7cm}}
    \toprule
    \textbf{Category} & \multicolumn{1}{l}{\textbf{Percentage}} & \textbf{Papers} \\
    \midrule
    \multicolumn{3}{l}{\textbf{Target phase}} \\
    \midrule
    Evasion attacks  & 87.5\% & \cite{grosse2017adversarial,li2019adversarial,chen2019android,podschwadt2019effectiveness,khoda2019selective,taheri2020adversarial,li2020adversarial,taheri2020defending,li2021framework,darwaish2021robustness,li2021robust,rathore2021robust,li2021can,zhao2021structural} \\
    Poisoning attacks & 12.5\% & \cite{li2021backdoor, severi2021explanation} \\
    \midrule
    \multicolumn{3}{l}{\textbf{Attacks scenarios}} \\
    \midrule
    White-box & 31.3\% & \cite{grosse2017adversarial,khoda2019selective,taheri2020adversarial,darwaish2021robustness,zhao2021structural} \\
    White-box \& Black-box & 37.5\% & \cite{podschwadt2019effectiveness,li2020adversarial,li2021framework,li2021robust,rathore2021robust,severi2021explanation} \\
    Black-box & 31.3\% & \cite{li2019adversarial,chen2019android,taheri2020defending,li2021can,li2021backdoor} \\
    \midrule
    \multicolumn{3}{l}{\textbf{Adversarial defenses strategies}} \\
    \midrule
    Specified & 75.0\% & \cite{grosse2017adversarial,chen2019android,podschwadt2019effectiveness,taheri2020adversarial,li2020adversarial,taheri2020defending,li2021framework,li2021robust,rathore2021robust,li2021can,li2021backdoor,zhao2021structural} \\
    Not Specified & 25.0\% & \cite{li2019adversarial,khoda2019selective,darwaish2021robustness,severi2021explanation} \\
    \bottomrule
    \end{tabular}%
  \label{tab:RQ3AdversarialAttacks}
\end{table}%

\subsubsection{Adversarial learning attacks and protections}
\label{sec:adversarial_trends}
Deep learning models are not resistant to adversarial attacks, which can cause a model to output an entirely incorrect prediction. 
Adversarial examples can be generated by just applying minor but intentional perturbations to original samples. 
A detailed taxonomy of adversarial sample crafting techniques and defensive techniques can be found in these survey work~\cite{papernot2016limitations,akhtar2018threat}. 
As discussed before, we discovered 16 primary studies (12\%) related to adversarial learning attacks and protections. 
Table~\ref{tab:RQ3AdversarialAttacks} presents a detailed summary for these 16 primary studies.

Focusing on the target phase of adversarial attacks, it is apparent that evasion attacks receive more research attention, accounting for 87.5\%.
Evasion attacks modify the data point at inference time, resulting in misclassification.
For example, Grosse et al.~\cite{grosse2017adversarial} perform adversarial evasion attacks on DNN-based malware detection models. 
This work exploited Jacobian based white-box attacks~\cite{papernot2016limitations} to generate adversarial examples, and the evaluation results indicated that the evasion algorithm can misclassify 63\% of all malware samples on Drebin.
Conversely, only two primary studies~\cite{li2021backdoor, severi2021explanation} investigated poisoning attacks, where the adversary's objective is to victimize the model training process.
It is interesting to observe that both of these two sources are devoted to backdoor poisoning attacks. 

Table~\ref{tab:RQ3AdversarialAttacks} shows that five sources focus on white-box attacks.
White-box attacks assume that the adversary has knowledge about the trained model such as model architectures and hyperparameters. 
For example, Grosse et al.~\cite{grosse2017adversarial} require the gradient information of DNN networks to craft adversarial examples.
Five sources perform black-box attacks on DL-based malware detection models, where the adversary requires no knowledge about the target classifier.
Six primary studies also assess the robustness of DNN models in both black-box and white-box scenarios.
These studies demonstrated that the attackers are slightly more vulnerable when having a limited knowledge of the model architecture.

With regard to adversarial defense strategies, Table~\ref{tab:RQ3AdversarialAttacks} indicates that 75\% of sources adopt defense mechanisms for adversarial attacks.
It is worth mentioning that most studies apply adversarial training and ensemble learning to defend against adversarial attacks~\cite{grosse2017adversarial, xiao2019android, podschwadt2019effectiveness, li2020adversarial, li2021framework, zhao2021structural}.

\begin{table}[t]
\scriptsize
  \centering
  \caption{A summary of DL-based malware evolution detection and protection approaches}
    \begin{tabular}{lcp{8em}p{16em}p{15em}}
    \toprule
    \multicolumn{1}{c}{\textbf{Tool}} & \textbf{Year} & \multicolumn{1}{c}{\textbf{DNN models}} & \multicolumn{1}{c}{\textbf{Approach}} & \multicolumn{1}{c}{\textbf{Model updating}} \\
    \midrule
    TESSERACT~\cite{pendlebury2019tesseract} & 2019  & MLP   & Proposing a new metric for time decay & Retraining without drift understanding, active learning, classification with rejection \\
    EveDroid~\cite{lei2019evedroid} & 2019  & MLP   & API semantics & - \\
    API-GRATH~\cite{zhang2020enhancing} & 2020  & MLP   & API semantics & Active learning \\
    Dr.Droid~\cite{fan2021heterogeneous} & 2021  & Transformers, GNN & Semantic relations; Heterogeneous temporal graph & -\\
    Li et al.~\cite{li2021can} & 2021  & MLP, CNN, RNN & Uncertainty & - \\
    SDAC~\cite{xu2020sdac} & 2020  & DNN   & API semantics & - \\
    CADE~\cite{yang2021cade} & 2021  & AE    & Concept drift detection & Retraining with drift understanding \\
    \bottomrule
    \end{tabular}%
  \label{tab:malwareEvolution}%
\end{table}%

\subsubsection{Malware evolution detection and defense}
\label{sec:malware_evolution_trend}
In the security domain, malware evolution has several similar concepts such as concept drift~\cite{yang2021cade}, time decay~\cite{pendlebury2019tesseract} and model aging~\cite{zhang2020enhancing}.
Fig.~\ref{fig:RQ1trend} shows that seven recent sources for detecting and defending malware evolution have been found.
In order to better understand the current research state, Table~\ref{tab:malwareEvolution} provides a summary of these seven primary studies.
It is worth highlighting that four primary studies~\cite{lei2019evedroid, zhang2020enhancing, xu2020sdac, fan2021heterogeneous} attempt to capture features' semantic similarity to slow down model aging.
Pendlebury et al.~\cite{pendlebury2019tesseract} propose a time-aware performance metric for measuring classifiers' resilience to malware evolution.
Indeed, these approaches merely slow down model performance degradation caused by malware evolution.
Thus, model updating approaches like model retraining or active learning are also frequently investigated to reverse and improve obsolete models~\cite{pendlebury2019tesseract, zhang2020enhancing}.
However, such model updating approaches remain evolution-insensitive, requiring periodical retraining.
In addition, this process often needs significant amounts of effort in labeling new samples.
To this end, Yang et al. employ contrastive learning to identify and understand drift malware samples before updating the aging models. 

\begin{tcolorbox}
\textbf{RQ3 What are the emerging and potential research trends?} 
\begin{itemize}
    \item Although there exists much research on DL-based on Android malware defenses, this topic still requires more in-depth analysis. 
    \item Malware evolution and adversarial attacks are two recent hot topics.
    \item How to improve the reliability, robustness and practicability of DL-based Android malware defense frameworks is a future challenge.
\end{itemize} 
\end{tcolorbox}


\section{Open issues and future trends}
In this work, we summarized the relevant sources of DL-based Android malware defenses and discussed research trends and challenges from various aspects in Section~\ref{sec: resultsanalysis}. 
Here we draw on these findings of this systematic review to provide a set of discussion points around the research and practice of Android security for future researchers.

\textbf{Android malware defenses remain a hot topic for further investigation.}
As our systematic review revealed, much research has been devoted to DL-based Android malware defenses in recent years, and the amount of relevant research is constantly increasing. 
It suggests that mobile security is a major concern nowadays. 
Mobile phones have become an integral part of people's daily life, and mobile users also pay much more attention to private mobile security, particularly the prevention of malicious applications.

However, the majority of existing research focuses on malware detection as a binary classification problem, which is far from enough to address current issues and improve mobile security. 
Malware remains among the most effective threats in the cyber space, and malware writers continue to update malware techniques to bypass security detection. 
As a result, this research requires more in-depth analysis rather than simply seeking a binary label. 
Other research aspects, like malware attribution/behaviors, malware variants, malware triage and treatments of infection, still receive scant attention.

\textbf{Data imbalance.} As shown in Section~\ref{sec:dataset}, the sources prefer to construct a relatively balanced dataset for performance evaluation purposes.
However, Android malware defenses suffer from serious sample sparsity and imbalance issues.
In the Android landscape, the number of goodware is significantly greater than the number of malware~\cite{pendlebury2019tesseract}.
Additionally, malware families are also highly imbalanced (thousands of samples in some families but only a few in others).
Many previous studies have demonstrated that imbalanced data distributions hinder the performance \cite{pendlebury2019tesseract, bai2020unsuccessful}.
Thus, how to develop effective solutions against data imbalance in DL-based Android malware defenses has gained immense research interest.
From our reviewed results, we identified two related studies to handle the data imbalance issue, including a Siamese network-based approach for imbalanced family classification~\cite{bai2020unsuccessful} and a BERT-based approach for imbalanced malware detection~\cite{oak2019malware}. 
However, these two studies are limited to relatively simple scenarios, and thus more efforts are still required to overcome the negative influences of data imbalance on Android malware defenses.

\textbf{Improving practicality and reliability is a priority.}
Our review also revealed that improving the practicality and reliability of DL-based Android malware defense approaches gathered increasing research interests. 
Although advanced deep learning techniques have been demonstrated to be effective at defending against malware attacks in a series of research experiments, how to effectively apply these approaches in practice remains unsolved. 
Future research should not only find a solution to overcome the limitation of mobile computing resources to deploy DL-based malware defense architectures, but also propose a comprehensive framework to tackle many realistic challenges such as internet privacy protection and information updating, which may require knowledge in other specific areas of cyber security and computer internet. 
Except for that, the black-box nature of deep neural networks poses a serious barrier to implementing these proposed approaches in practice. 
How to improve the transparency of the malware defense process will be a noteworthy research topic in the future.

\textbf{Deep learning in Android security is still in an early stage.}
Compared with other research domains, research towards deep learning in the Android security community seems to be relative singleness. 
First, supervised DNN networks are the most investigated, and these studies usually consider DNN networks with three to four layers in their proposed architectures. 
Secondly, more advanced deep learning approaches like reinforcement learning and online learning are covered by a minority of papers.
Thirdly, most previous works adopt deep learning techniques on some simple tasks like binary Android malware classification.
In fact, deep learning has made considerable achievements in computer vision and natural language processing.
These advanced techniques have been demonstrated a powerful capability in solving many complex tasks~\cite{lecun2015deep}.
For example, Wu et al.~\cite{wu2020android} introduce an attention mechanism to improve the interpretability of Android malware detection models.
Thus, it is promising to apply advanced deep learning techniques to assist us in solving more complex and specific issues in Android security.
At the same time, while supervised learning is dominant in the Android security domain, labeling data is time-consuming and requires substantial expertise.
As described by deep learning textbook by Lecun et al.~\cite{lecun2015deep}, unsupervised learning belongs to the future of deep learning.
We encourage our fellow researchers to make more efforts on unsupervised deep learning in Android security to make more advancements in the future.

\textbf{APK embedding is an important but untouched topic.}
Different from image or word information, Android apps are composed of multiple complex data.
Android APK is a zip archive consisting of multiple files. 
Through reverse engineering, various types of features like permissions and opcode can be extracted for further analysis. 
In fact, deep learning still struggles to model these complex data modalities \cite{pouyanfar2018survey}. 
Therefore, existing research either transforms APK into a single type of feature or designs a multimodel deep learning architecture to handle them. 
To obtain a formalized representation compatible with DL models, embedding techniques from CV and NLP are introduced to encode features, but these techniques may be relatively shallow for APK files with complicated structures. 
As for Android security, there is still a long way to explore "APK embedding" for DNN models.

\section{Threats to Validity}
Even though this systematic review was conducted by following a well-established methodology \cite{kitchenham2004procedures}, we can't guarantee that our results covered all relevant studies, caused by some limitations during the review process. 
Thus, this section describes possible threats to the validity of our empirical study. 

\emph{Search items and strategies}. One main potential threat is relevant publication collection bias. 
In order to locate relevant studies, we described a list of search strings and search databases in Section \ref{sec:searchstrategy}. 
Search strings were formulated by different items from both software engineering and artificial intelligence domains. 
Although we added alternative spellings and synonyms for search items, we may still miss some search items. 
For instance, deep learning is a rapidly developing research field and AI scientists would continue to propose new DL techniques in a short time; thus, identifying all relevant DL items is a challenge. 
To minimize this issue, we retained the DL items described in \cite{lecun2015deep,goodfellow2016deep, SCHMIDHUBER201585} surveyed by AI experts. 
After search strings were identified, five well-known electronic databases were used to collect relevant studies. 
We also conducted further searching processes on two popular research citation engines and a backward snowballing to cover the relevant publications in the broadest sense.  

\emph{Data selection bias}. The selection of publications was conducted by one researcher only, which may cause missing studies. 
Despite that, all authors formulated an appropriate study selection scheme together, and the other three authors gave effective and detailed feedback and monitored review execution closely throughout the review process. 
On the other hand, in order to reduce the impact of subjective factors in the quality evaluation process, H5-index of the venues was introduced as a quality appraisal criterion. 
Although the H5-value may change over time, the influential papers from the top venues were guaranteed to be taken into account.

\section{Conclusions}
This article provides an in-depth examination of the use of deep learning in Android malware defenses.
Additionally, the study discusses research objectives, characteristics, approaches, and challenges associated with Android malware defenses using deep learning. 
We collected 132 relevant studies.
Our reviewed results indicate that deep learning techniques are becoming a powerful and promising tool to defend against Android malicious applications. 
We discovered that (1) most studies are conducted to detect malware, but other types of more detailed analysis on malicious apps are receiving increasing attention; 
(2) static program analysis is widely used to collect features, and semantic features are frequently occurring; 
(3) various DNN architectures are employed to analyze malware, among which MLPs and CNNs are the most widely used; 
(4) most approaches are performed as a supervised classification task; 
(5) distributed analysis and on-device analysis is gradually valued;
(6) adversarial learning and malware evolution are two recent hot topics.


\bibliographystyle{sample-format/ACM-Reference-Format}
\bibliography{sample-base}

\end{document}